\documentclass[ejs,preprint]{imsart}

\arxiv{1406.1780}
\usepackage{graphicx}
\usepackage{algpseudocode}
\usepackage{algorithm}
\usepackage{subfigure}
\usepackage{comment}

\usepackage{amsmath, amsthm, amssymb}
\RequirePackage[round]{natbib}

\RequirePackage[colorlinks,citecolor=blue,urlcolor=blue]{hyperref}

\usepackage[usenames,dvipsnames,svgnames,table]{xcolor}

\newtheorem{thm}{Theorem}

\let\hat\widehat

\newcommand\R{\mathbb{R}}

\newcommand\norm[1]{\|#1\|}
\newcommand\MISE{{\sf MISE}}
\newcommand\Diag{{\sf Diag}}
\newcommand\MDS{{\sf MDS}}
\newcommand\Haus{{\sf Haus}}

\newcommand\cC{{\cal C}}
\newcommand\cL{{\cal L}}
\newcommand\cM{{\cal M}}

\DeclareMathOperator{\dest}{dest}

\newenvironment{enum}{
\begin{enumerate}
  \setlength{\itemsep}{1pt}
  \setlength{\parskip}{0pt}
  \setlength{\parsep}{0pt}
}{\end{enumerate}}

\catcode`@=11
\newskip\beforeproofvskip
\newskip\afterproofvskip
\beforeproofvskip=\medskipamount
\afterproofvskip=\bigskipamount

\def\prooftag{Proof}
\def\proofskip{\enspace}

\def\proof{\@ifnextchar[{\@@proof}{\@proof}}  
\def\@startproof{\par\vskip\beforeproofvskip\leavevmode}
\def\@proof{\@startproof{\scshape\prooftag.}\proofskip}
\def\@@proof[#1]{\@startproof {\scshape\prooftag #1.}\proofskip}

\catcode`@=12

\begin{document}
\begin{frontmatter}

\title{A Comprehensive Approach to Mode Clustering}
\runtitle{Mode Clustering}

\begin{aug}
  \author{Yen-Chi Chen\ead[label=e1]{yenchic@andrew.cmu.edu}},
  \and
  \author{Christopher R. Genovese\ead[label=e2]{genovese@stat.cmu.edu}},
  \and 
  \author{Larry Wasserman\ead[label=e3]{larry@stat.cmu.edu}}
  
  \address{Carnegie Mellon University,\\
  5000 Forbes Avenue,\\ 
  Pittsburgh, PA 15213 \\
           \printead{e1,e2,e3}}


  \runauthor{Chen et al.}

\end{aug}

\begin{abstract}
Mode clustering is a nonparametric method for clustering
that defines clusters using the
basins of attraction of a density estimator's modes.
We provide several enhancements to mode clustering:
(i)~a soft variant of cluster assignment, 
(ii)~a measure of connectivity between clusters,
(iii)~a technique for choosing the bandwidth,
(iv)~a method for denoising small clusters, and
(v)~an approach to visualizing the clusters.
Combining all these enhancements gives us a 
complete procedure for clustering 
in multivariate problems.
We also compare mode clustering to other clustering methods in several examples.
\end{abstract}




\end{frontmatter}





\section{Introduction}

Mode clustering is a nonparametric clustering method
\citep{Azzalini:2007la, cheng1995mean,chazal2011persistence, Comaniciu,Fukunaga,li2007nonparametric, Chacon2012, Arias-Castro2013, Chacon2014}
with three steps:
(i)~estimate the density function,
(ii)~find the modes of the estimator,
and (iii)~define clusters by the basins of attraction of these modes.

There are several advantages to using mode clustering relative to other commonly-used methods:
\begin{enum}
\item There is a clear population quantity being estimated.
\item Computation is simple:
the density can be estimated
with a kernel density estimator, and the modes and basins of attraction
can be found with the mean-shift algorithm.
\item There is a single tuning parameter to choose, namely, the bandwidth of the 
density estimator.
\item It has strong theoretical support since it depends only on density estimation
and mode estimation
\citep{Arias-Castro2013,romano1988weak,romano1988bootstrapping}.
\end{enum}

\begin{figure}
\center
\includegraphics[width=2 in, height=2 in]{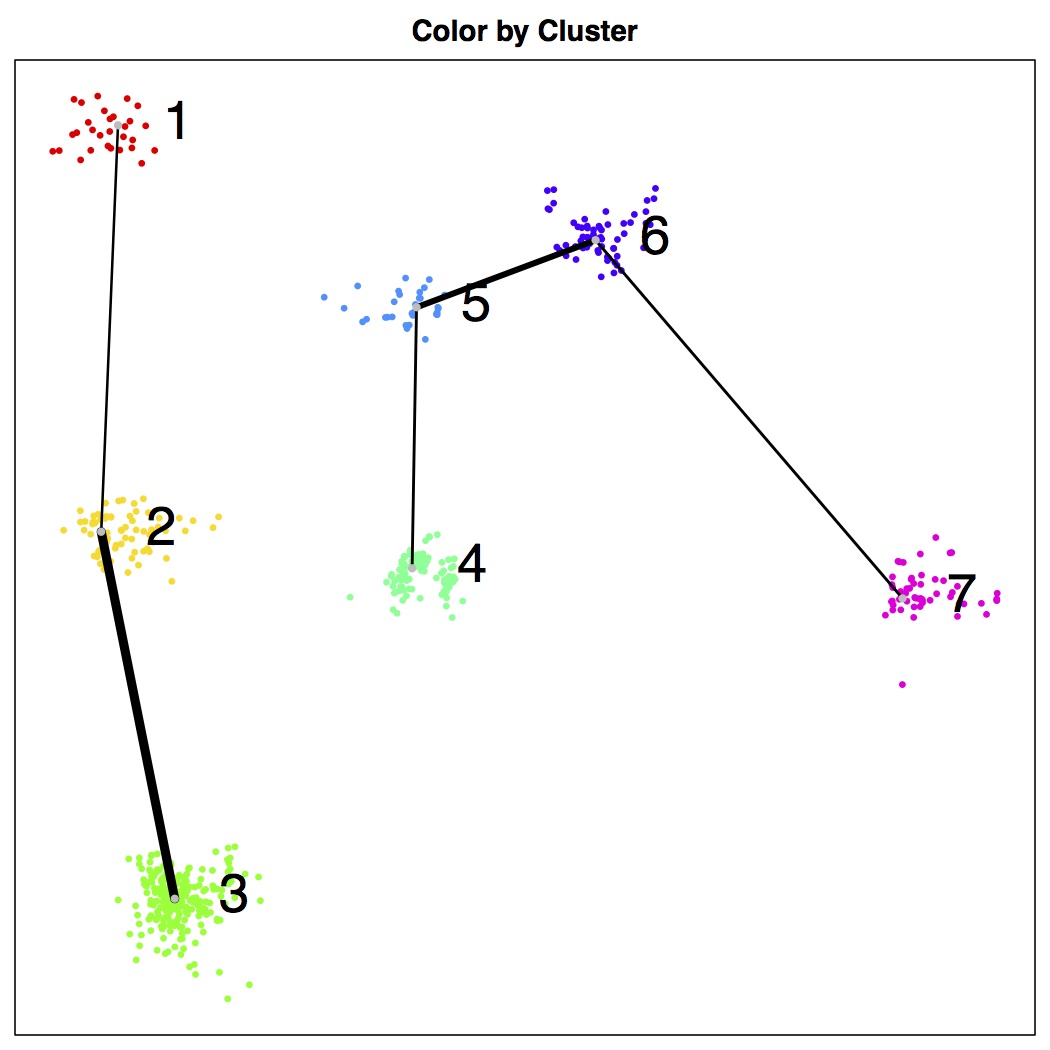}
\caption{An example for visualizing multivariate mode clustering. 
This is the Olive Oil data, which has dimension $d=8$.
Using the proposed methods in this paper, we identify $7$ clusters
and the connections among clusters are represented by edges 
(width of edge shows the strength of connection). 
More details can be found in section \ref{sec::OO}.}
\label{Fig::EX0}
\end{figure}

Despite these advantages, there is room for improvement.
First, mode clustering results is a hard assignment;
there is no measure of uncertainty as to how well-clustered a data point is.
Second, it is not clear how to visualize the clusters when the dimension is greater than two.
Third, one needs to choose the bandwidth of the kernel estimator.
Fourth, in high dimensions, mode clustering tends to produce 
tiny clusters which we call ``clustering noise.''
In this paper, we propose solutions to all these issues
which leads to a complete, comprehensive approach to model clustering.
Figure~\ref{Fig::EX0} shows an example of mode clustering
for a multivariate data with our visualization method ($d=8$).

\vspace{1cm}

\emph{Related Work}. Mode clustering is based on the mean-shift
algorithm \citep{Fukunaga,cheng1995mean,Comaniciu} which is a popular
technique in image segmentation.
\cite{li2007nonparametric, Azzalini:2007la} formally introduced mode
clustering to the statistics literature. The related idea of clustering based on
high density regions was proposed in \cite{Hartigan1975}.
\cite{Chacon2011, Chacon2012} propose several methods for selecting the
bandwidth for estimating the derivatives of the
density estimator which can in turn be used as a bandwidth selection
rule for mode clustering.
The idea of merging insignificant modes is related to
the work in \cite{li2007nonparametric, Fasy2014, chazal2011persistence,
Chaudhuri2010, kpotufe2011pruning}.

\vspace{1cm}

\emph{Outline}.
In Section \ref{sec::MC}, we review the basic idea of mode clustering.
In Section \ref{sec::SC}, we discuss soft cluster assignment methods.
In Section \ref{sec::connect}, we define a measure of connectivity among clusters 
and propose an estimate of this measure.
In Section \ref{sec::consistency} we prove consistency of the method.
In Section \ref{sec::h}, we describe a rule for bandwidth selection in mode clustering.
Section \ref{sec::merge} deals with the problems of tiny clusters which occurs
more frequently as the dimension grows.
In Section \ref{sec::V}, we introduce a visualization technique for high-dimensional data
based on multidimensional scaling.
We provide several examples in section \ref{sec::ex}. 
The R-code for our approaches can be found in 
\url{http://www.stat.cmu.edu/~yenchic/EMC.zip}.

\section{Review of Mode Clustering}	\label{sec::MC}

\begin{figure}
\centering
	\subfigure[Attraction basins]
	{
		\includegraphics[width=2 in, height=2 in]{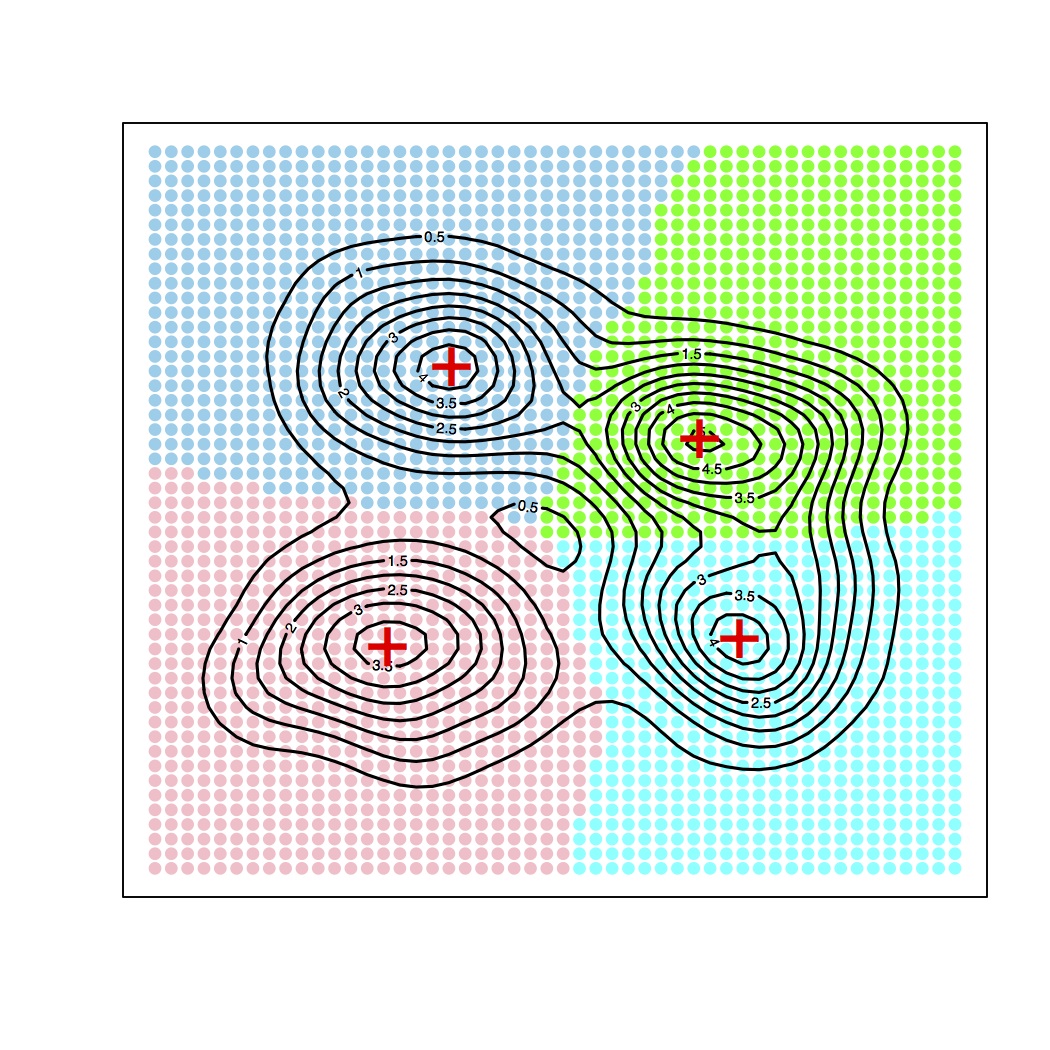}
	}
	\subfigure[Mean shift]
	{
		\includegraphics[width=2 in, height=2 in]{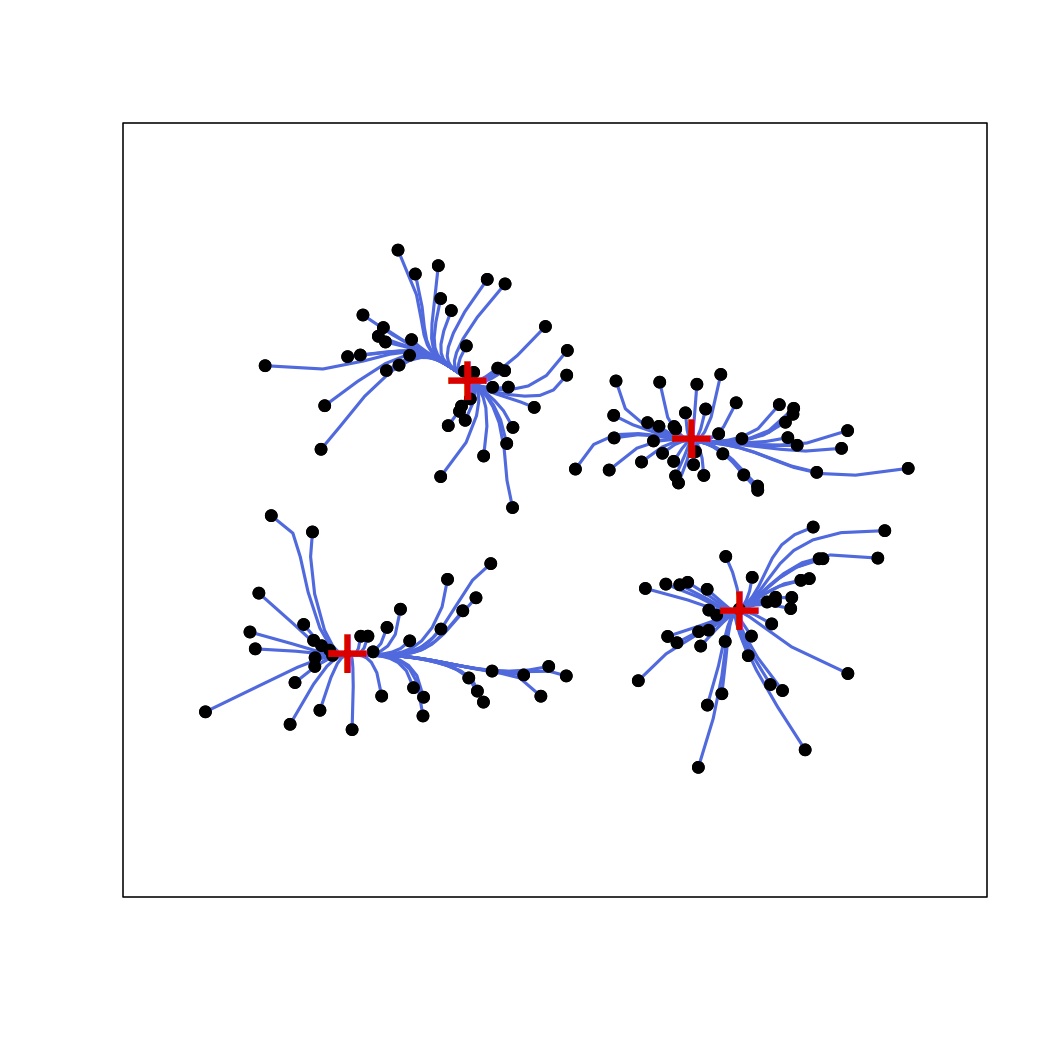} 
	}
\caption{The mode clustering. (a): the attraction basins for each mode
given a smooth function. (b): the mean shift algorithm to cluster
data points. The red crosses are the local modes.}
\label{Fig::EX}
\end{figure}


Let $p$ be the density function of a random vector $X\in\R^d$.
Throughout the paper, we assume $p$
has compact support $\mathbb{K}\subset \R^d$. 
Assume
that $p$ has $k$ local maxima $\cM = \{m_1,\cdots,m_k\}$ and is a
Morse function \citep{morse1925relations, morse1930foundations,
banyaga2004lectures}, meaning that the Hessian of $p$ at each critical
point is non-degenerate.  We do not assume that $k$ is known.  Given
any $x\in\R^d$, there is a unique gradient ascent path starting at $x$
that eventually arrives at one of the modes
(except for a set of $x$'s of measure 0).
We define the clusters as
the `basins of attraction' of the modes
\citep{chacon2012clusters}, i.e., the sets of points
whose ascent paths have the same mode.  Now we give more detail.

An \emph{integral curve} through $x$ is a path $\pi_x:
\R\mapsto\R^d$ such that $\pi_x(0)=x$ and
\begin{equation}
\pi_x'(t) = \nabla p(\pi_x(t)).
\end{equation}
A standard result
in Morse theory is that integral curves never intersect except at critical points,
so the curves partition the space
\citep{morse1925relations, morse1930foundations,
banyaga2004lectures}.
We define the destination for the integral curve starting at $x$ by
\begin{equation}
\dest(x) = \underset{t\rightarrow \infty}{\lim} \pi_x(t).
\end{equation}
Then
$\dest(x)=m_j$ for some mode $m_j$ for all $x$
except on a set $E$ with Lebesgue measure $0$ ($E$ contains points that
are on the boundaries of clusters and whose paths
lead to saddle points).
For each mode $m_j$ we define
the \emph{basin of attraction} of $m_j$ by
\begin{equation}
C_j =\{x: \dest(x) = m_j\},\quad j=1,\cdots,k.
\end{equation}
$C_j$ is also
called the \emph{ascending manifold} \citep{Guest2001}
or the \emph{stable manifold} \citep{morse1925relations, morse1930foundations,
banyaga2004lectures}.
The partition $\cC = \{C_1,\ldots, C_k\}$ is called the \emph{Morse complex} of $p$.
These are the population clusters.

In practice, $p(x)$ is unknown and we need to estimate it.
A common way to do this is via the kernel density estimator (KDE).
Let $X_1,\ldots, X_n$ be a random sample from $p$,
and let $K$ be a smooth, symmetric kernel.
The KDE with bandwidth $h > 0$ is defined by
\begin{equation}
\hat{p}_h(x) = \frac{1}{nh^d} \sum_i K\left(\frac{||x-X_i||}{h}\right).
\end{equation}
The modes
$\hat\cM = \{\hat m_1,\ldots, \hat m_{\hat k}\}$
of $\hat p_n$ 
and the integral-curve destinations under $\hat p_n$ of any point $x$,
$\hat\dest(x)$,
are both easily found using the
\emph{mean-shift} algorithm
\citep{Fukunaga,cheng1995mean,Comaniciu}.
The corresponding basins of attraction
are
\begin{align} 
\hat{C}_j &= \{x\in\R^d:\; \hat\dest(x) =\hat{m}_j\}, \quad j=1,\cdots,{\hat k} \\
\hat\cC &= \{\hat C_1,\ldots, \hat C_{\hat k}\}
\end{align} 
and the sample clusters are defined by
\begin{equation}
\mathcal{X}_j = \{ X_i:\ X_i \in \hat C_j \} = 
\{X_i: \hat\dest(X_i)=\hat{m}_j\}.
\label{eq::MC3}
\end{equation}

\section{Soft Clustering}	\label{sec::SC}
Mode clustering is a type of hard clustering,
where each observation is assigned to one and only one cluster.
Soft clustering methods \citep{mclachlan2004finite,Lingras2002, Nock2006,Peters2013}
attempt to capture the uncertainty in this assignment.
This is typically represented by an assignment vector for each point
that is a probability distribution over the clusters.
For example, whereas a hard-clustering method might assign a point $x$
to cluster 2, a soft clustering might give $x$ an assignment vector
$a(x) = (0.01, 0.8, 0.01, 0.08, 0.1)$,
reflecting both the high confidence that $x$ belongs to cluster $2$ 
the nontrivial possibility that it belongs to cluster $5$.

Soft clustering can capture two types of
cluster uncertainty: population level (intrinsic difficulty) and
sample level (variability). The population level uncertainty
originates from the fact that even if $p$ is known,
some points are more strongly related to their modes than others.
Specifically, for a point $x$ near the boundaries between
two clusters, say $C_1,C_2$, the associated soft assignment vector
$a(x)$ should have $a_1(x)\approx a_2(x)$.
The sample level
uncertainty comes from the fact that 
$p$ has been estimated by $\hat p$.
The soft
assignment vector $a(x)$ is designed to capture both types of
uncertainty.

\vspace{1cm}

{\bf Remark:}
The most common soft-clustering method is to use a mixture model.
In this approach, we represent cluster membership by a latent variable
and use the estimated distribution of that latent variable as the assignment vector.
In the appendix we discuss mixture-based soft clustering.

\subsection{Soft Mode Clustering}

One way to obtain soft mode clustering
is to use a distance from a given point $x$ to all the local modes.
The idea is simple: if $x$ is close to a mode $m_j$,
the soft assignment vector should have a higher $a_j(x)$.
However, converting a distance to a soft assignment vector 
involves choosing some tuning parameters.

Instead, we now present a more direct method based on a diffusion 
that does not require any conversion of distance.
Consider starting a diffusion at $x$.
We define the soft clustering as the probability that the diffusion 
starting at $x$ leads to a particular mode,
before hitting any other mode.
That is, let $a^{HP}(x) = (a^{HP}_1(x),\ldots, a^{HP}_k(x))$
where $a^{HP}_j(x)$ is the conditional probability that mode $j$ is the first mode
reached by the diffusion, given that it reaches one of the modes.
In this case $a(x)$ is a probability vector and so is easy to interpret.

In more detail, let 
$$
K_{h}(x,y) = K\left(\frac{\norm{x-y}}{h}\right).
$$
Then
$$
q_{h}(y|x)= \frac{K_{h}(x,y)p(y)}{\int K_{h}(x,y)dP(y)},
$$
defines a Markov process
with $q_h(y|x)$ being the probability of jumping to $y$ given that the process is at $x$.
Fortunately, we do not actually have to run the diffusion to estimate $a^{HP}(x)$.

An approximation to the above diffusion process restricted to $x,y$ in
$\{\hat{m}_1,\ldots,\hat{m}_{\hat{k}}, X_1,\ldots,X_n\}$ is as
follows. We define a Markov chain that has $\hat{k}+n$ 
states. The first $\hat{k}$ states are the estimated local modes
$\hat{m}_1,\ldots, \hat{m}_{\hat{k}}$ and are absorbing states.
That is, the Markov process stops when it hits any of the first $\hat{k}$ state. 
The other $n$ states correspond to the data points $X_1,\ldots,X_n$. The transition
probability from each $X_i$ is given by
\begin{equation}
\begin{aligned}
\mathbf{P}(X_i\rightarrow \hat{m}_l) &= \frac{K_h(X_i,\hat{m}_j)}{\sum_{j=1}^n K_h(X_i,X_j)+\sum_{l=1}^{\hat{k}} K_h(X_i,\hat{m}_l)}\\
\mathbf{P}(X_i\rightarrow X_j) &= \frac{K_h(X_i,X_j)}{\sum_{j=1}^n K_h(X_i,X_j)+\sum_{l=1}^{\hat{k}} K_h(X_i,\hat{m}_l)}
\end{aligned}
\end{equation}
for $i,j=1,\ldots,n$ and $l=1,\ldots,\hat{k}$. Thus, the transition matrix $\mathbf{P}$ is
\begin{equation}
\mathbf{P} = \begin{bmatrix}
	\mathbf{I}& 0\\
	S&T
	\end{bmatrix},
\end{equation}
where $\mathbf{I}$ is the identity matrix and $S$ is an $n\times
\hat{k}$ matrix with element $S_{ij}= \mathbf{P}(X_i\rightarrow
\hat{m}_j)$ and $T$ is an $n\times n$ matrix with element $T_{ij} =
\mathbf{P}(X_i\rightarrow X_j)$. Then by Markov chain theory, the
absorbing probability from $X_i$ onto $\hat{m}_j$ is given by
$\hat{A}_{ij}$ where $\hat{A}_{ij}$ is the $(i,j)$-th element of the
matrix
\begin{equation}
\hat{A} = S(\mathbf{I}-T)^{-1}.
\end{equation}
We define the soft assignment vector by $\hat{a}^{HP}_j(X_i) = \hat{A}_{ij}$.

\section{Measuring Cluster Connectivity} \label{sec::connect}

In this section, we propose a technique that uses the soft-assignment
vector to measure the connectivity among clusters.  Note that the
clusters here are generated by the usual (hard) mode clustering.

Let $p$ be the density function and $C_1,\ldots,C_k$ be the clusters corresponding to 
the local modes $m_1,\ldots,m_k$.
For a given soft assignment vector $a(x): \R^d \mapsto \R^k$,
we define the \emph{connectivity} of cluster $i$ and cluster $j$ by
\begin{equation}
\begin{aligned}
\Omega_{ij} &= \frac{1}{2}\Bigl(\mathbb{E}\bigl(a_i(X)|X\in C_j\bigr)+\mathbb{E}\bigl(a_j(X)|X\in C_i\bigr)\Bigr)\\
           &= \frac{1}{2}\frac{\int_{C_i} a_j(x)p (x)dx}{\int_{C_i}p(x)dx} +\frac{1}{2}\frac{\int_{C_j} a_i(x)p (x)dx}{\int_{C_j}p(x)dx}.
\end{aligned}
\label{eq::CC1}
\end{equation}
Each $\Omega_{ij}$ is a population level quantity that depends only on how we determine the soft assignment vector.
Connectivity will be large when two clusters are close and the boundary between them has high density. 
If we think of the (hard) cluster assignments as class labels,
connectivity is analogous to the mis-classification rate between class $i$ and class $j$. 

An estimator of $\Omega_{ij}$ is 
\begin{equation}
\hat{\Omega}_{ij}  = \frac{1}{2}\Bigl(\frac{1}{N_i} \sum_{l=1}^n \hat{a}_j(X_l) 1(X_l\in\hat{C}_i) +\frac{1}{N_j} \sum_{l=1}^n \hat{a}_i(X_l) 1(X_l\in\hat{C}_j)\Bigr),\quad i,j=1,\ldots, \hat{k},
\label{eq::CC2}
\end{equation}
where $N_i =\sum_{l=1}^n 1(X_l\in\hat{C}_i)$ is the number of sample in cluster $\hat{C}_i$. 
Note that when $n$ is sufficiently large, each estimated mode is a consistent estimator to one population mode \citep{chazal2014robust}
but the ordering might be different. For instance, the first estimated mode $\hat{m}_1$ might be the 
estimator for the third population mode $m_3$.
After relabeling, we can match the ordering of both population and estimated modes.
Thus, after permutation of columns and rows of $\hat{\Omega}$, $\hat{\Omega}$ will be a consistent estimator to $\Omega$.
The matrix $\hat{\Omega}$ is a summary statistics for the connectivity between clusters.
We call $\hat{\Omega}$ the matrix of connectivity or the connectivity matrix.

The matrix $\hat{\Omega}$ is useful as a dimension-free, summary-statistic to describe 
the degree of overlap/interaction among the clusters,
which is hard to observe directly when $d > 2$.
Later we will use $\hat{\Omega}$ to describe the relations among clusters
while visualizing the data.

\section{Consistency}
\label{sec::consistency}

Local modes play a key role in mode clustering.
Here we discuss the consistency of mode estimation.
Despite the fact that the consistency for estimating a global mode
has been established
\citep{romano1988bootstrapping, romano1988weak, Pollard1985, Arias-Castro2013, Chacon2014,chazal2014robust, chen2014nonparametric}, 
there is less work on
estimating local modes .

Here we adapt the result in \cite{Chen2014GMRE}
to describe the consistency of estimating local modes
in terms of the Hausdorff distance.
For two sets $A,B$,
the Hausdorff distance is 
\begin{equation}
\Haus(A,B) = \inf\{r: A\subset B\oplus r, B\subset A\oplus r\},
\end{equation}
where $A\oplus r = \{y: \min_{x\in A}\norm{x-y}\leq r\}$.
The Hausdorff distance is a generalized $L_\infty$ metric for sets.

Let $K^{(\alpha)}$ be the $\alpha$-th derivative of $K$ and $\mathbf{BC}^r$
denotes the collection of functions with bounded continuously derivatives up
to the $r$-th order.
We consider the following two common assumptions on kernel function:
\begin{itemize}
\item[(K1)] The kernel function $K\in\mathbf{BC}^3$ and is symmetric, non-negative and 
$$\int x^2K^{(\alpha)}(x)dx<\infty,\qquad \int \left(K^{(\alpha)}(x)\right)^2dx<\infty
$$ 
for all $\alpha=0,1,2,3$.
\item[(K2)] The kernel function satisfies condition $K_1$ of
  \cite{Gine2002}. That is, there exists some $A,v>0$ such that for
  all $0<\epsilon<1$,
$\sup_Q N(\mathcal{K}, L_2(Q), C_K\epsilon)\leq \left(\frac{A}{\epsilon}\right)^v,$
where $N(T,d,\epsilon)$ is the $\epsilon-$covering number for a
semi-metric space $(T,d)$ and
$$
\mathcal{K} = 
\Biggl\{u\mapsto K^{(\alpha)}\left(\frac{x-u}{h}\right)
: x\in\R^d, h>0,|\alpha|=0,1,2,3\Biggr\}.
$$
\end{itemize}

The assumption (K1) is a smoothness condition on the kernel function.
(K2) controls the complexity of the kernel function
and is used in
\citep{Gine2002, Einmahl2005,Genovese2012a,Arias-Castro2013, Chen2014}.

\begin{thm}
[Consistency of Estimating Local Modes]
\label{thm::LM}
Assume $p\in \mathbf{BC}^3$ and the kernel function $K$ satisfies (K1-2).
Let $C_3$ be the bound for the partial derivatives of $p$ up to the third order
and $\hat{\mathcal{M}}_n\equiv\hat{\mathcal{M}}$ be the collection
of local modes of the KDE $\hat{p}_n$ and $\mathcal{M}$ be the local modes
of $p$. Let $\hat{K}_n$ be the number of estimated local modes and $K$ be
the number of true local modes.
Assume
\begin{itemize}
\item[(M1)] There exists $\lambda_*>0$ such that
$$
0<\lambda_*\leq |\lambda_1(m_j)|, \quad j=1,\cdots, k,
$$
where $\lambda_1(x)\leq \cdots\leq \lambda_d(x)$ are the eigenvalues of Hessian matrix of $p(x)$.

\item[(M2)] There exists $\eta_1>0$ such that
$$
\{x: \norm{\nabla p(x)}\leq \eta_1, 0>-\lambda_*/2\geq \lambda_1(x)\} \subset \mathcal{M}\oplus \frac{\lambda_*}{2dC_3},
$$
where $\lambda_*$ is defined in (M1).
\end{itemize}


Then when $h$ is sufficiently small and $n$ is sufficiently large,
\begin{itemize}
\item [1.] (Modal consistency) there exists some constants $A,C>0$ such that
$$
\mathbb{P}\left(\hat{k}_n\neq k\right)\leq Ae^{-Cnh^{d+4}};
$$
\item[2.] (Location convergence) the Hausdorff distance between local modes and their estimators satisfies
$$
\Haus\left(\hat{\mathcal{M}}_n,\mathcal{M}\right) = O(h^2)+O_P\left(\sqrt{\frac{1}{nh^{d+2}}}\right).
$$
\end{itemize}

\end{thm}

The proof is in appendix.
Actually, the assumption (M1) always hold whenever we assume $p$
to be a Morse function. We make it an assumption just for the convenience of the proof.
The second condition (M2) is a regularity on $p$
which requires that points with similar behavior
(near $0$ gradient and negative eigenvalues) to local modes
must be close to local modes.
Theorem~\ref{thm::LM} states two results:
consistency for estimating the number of local modes
and consistency for estimating the location of local modes.
An intuitive explanation for the first result
is from the fact that as long as the gradient and Hessian matrix of KDE $\hat{p}_n$
are sufficiently closed to the true gradient and Hessian matrix,
condition (M1, M2) guarantee the number of local modes is the same as truth.
Applying Talagrand's inequality \citep{Talagrand1996} 
we obtain exponential concentration which gives the desired result.
The second result follows from applying a Taylor expansion
of the gradient
around each local mode,
the difference between local modes and their estimators 
is proportional to the error in estimating the gradients.
The Hausdorff distance can be decomposed into bias $O(h^2)$
and variance $O_P\left(\sqrt{\frac{1}{nh^{d+2}}}\right)$.

\subsection{Bandwidth Selection}	
\label{sec::h}

A key problem in mode clustering is the choice of the smoothing bandwidth $h$.
Because mode clustering is based on the gradient of the density function,
we choose a bandwidth targeted at gradient estimation.
From standard non-parametric density estimation theory, the estimated gradient
and the true gradient differ by
\begin{equation}
\norm{\nabla \hat{p}_n(x) - \nabla p(x)}^2_2 = O(h^4)+ O_P\Bigl(\frac{1}{nh^{d+2}}\Bigr)
\end{equation}
assuming $p$ has two smooth derivatives,
see \cite{Chacon2011,Arias-Castro2013}.
In non-parametric literature, a common error measure is the mean integrated square error (MISE).
The MISE for the gradient is
\begin{equation}
\MISE (\nabla \hat{p}_n) = \mathbb{E}\left(\int \norm{\nabla \hat{p}_n(x)- \nabla p(x)}^2_2 dx\right) = O\left(h^4\right)+O\left(\frac{1}{nh^{d+2}}\right)
\label{eq::h}
\end{equation}
when we assume (K1); see Theorem 4 of \citep{Chacon2011}.
Thus, it follows that the asymptotically optimal bandwidth should be
\begin{equation}
h = Cn^{-\frac{1}{d+6}},
\end{equation}
for some constant $C$.
In practice, we do not know $C$, so we need a concrete rule to select it.
We recommend a normal reference rule (a slight modification of \cite{Chacon2011}):
\begin{equation}
h_{NR} = \bar{S}_n\times \Bigl(\frac{4}{d+4}\Bigr)^{\frac{1}{d+6}}n^{-\frac{1}{d+6}},
\qquad\bar{S}_n = \frac{1}{d}\sum_{j=1}^d S_{n,j}
\label{eq::h0}
\end{equation}
where $S_{n,j}$ is the sample standard deviation along $j$-th coordinate.
We use this for two reasons.
First, it is known that the normal reference rule tends to
oversmooth \citep{Sheather2004}, which is typically good for clustering.
And second, the normal reference rule is easy to compute even in high dimensions. 
Note that this normal reference rule is optimizing asymptotic MISE for 
multivariate Gaussian distirbution with covariance matrix $\sigma \mathbf{I}$.
Corollary 4 of \cite{Chacon2011} provides a formula for the general covariance matrix case.
In data analysis, it is very common to normalize the data first
and then perform mode clustering.
If we normalize the data, the reference rule \eqref{eq::h0}
reduces to 
$h_{NR} =  \Bigl(\frac{4}{d+4}\Bigr)^{\frac{1}{d+6}}n^{-\frac{1}{d+6}}$.
For a comprehensive survey on the bandwidth selection, we refer the readers to \cite{Chacon2012}.

In addition to the MISE, another common metric for measuring the quality
of the estimator $\nabla \hat{p}_n$ is the $\cL^{\infty}$ norm,
which is defined by
\begin{equation}
\norm{\nabla \hat{p}_n - \nabla p}_{\max, \infty} = \sup_x \norm{\nabla \hat{p}_n(x)-\nabla p(x)}_{\max},
\end{equation}
where $\norm{v}_{\max}$ is the maximal norm for a vector $v$.

The rate for $\cL^{\infty}$ is 
\begin{equation}
\norm{\nabla \hat{p}_n - \nabla p}_{\max, \infty} = O\left(h^2\right)+O_P\left(\sqrt{\frac{\log n}{nh^{d+2}}}\right)
\label{eq::sup}
\end{equation}
when we assume (K1--2) and $p\in\mathbf{BC}^3$ \citep{genovese2009path,Genovese2012a,Arias-Castro2013, Chen2014}.
This suggests selecting the bandwidth by
\begin{equation}
h = C'\left(\frac{\log n}{n}\right)^{\frac{1}{d+6}}.
\end{equation}
However, no general rule has been proposed based on this norm.
The main difficulty is that no analytical form for the big $O$ term has been found.

{\bf Remark.}
Comparing the assumptions in Theorem~\ref{thm::LM}, equations \eqref{eq::h} and \ref{eq::sup}
gives an interesting result:
If we assume $p\in\mathbf{BC}^3$ and (K1), we obtain
consistency in terms of the MISE. If further we assume (K2), we
get the consistency in terms of the supremum-norm.
Finally, if we have conditions (M1-2), we obtain mode consistency.


\section{Denoising Small Clusters}	\label{sec::merge}

In high dimensions, mode clustering tends to produce many small clusters ,
that is, clusters with few data points.
We call these small clusters, \emph{clustering noise}.
In high dimensions, the variance creates small bumps
in the KDE which then creates clustering noise.
The emergence of clustering noise is consistent with Theorem~\ref{thm::LM};
the convergence rate is much slower when $d$ is high.

\begin{figure}
\centering
	\subfigure[Scatter plot for first two coordinates]
	{
		\includegraphics[width=2 in, height=2 in]{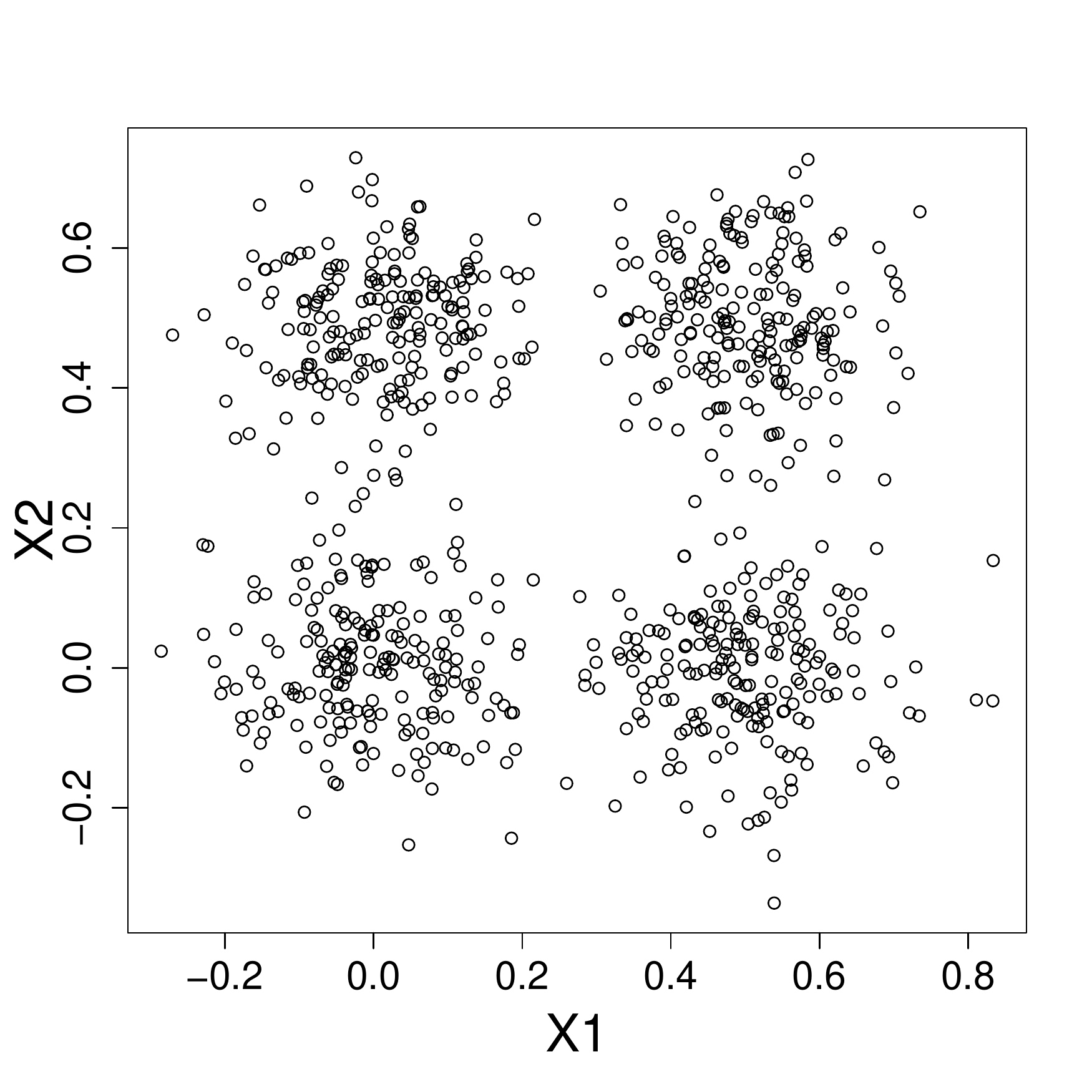}
	}
	\subfigure[Ordered size of clusters]
	{
		\includegraphics[width=2 in, height=2 in]{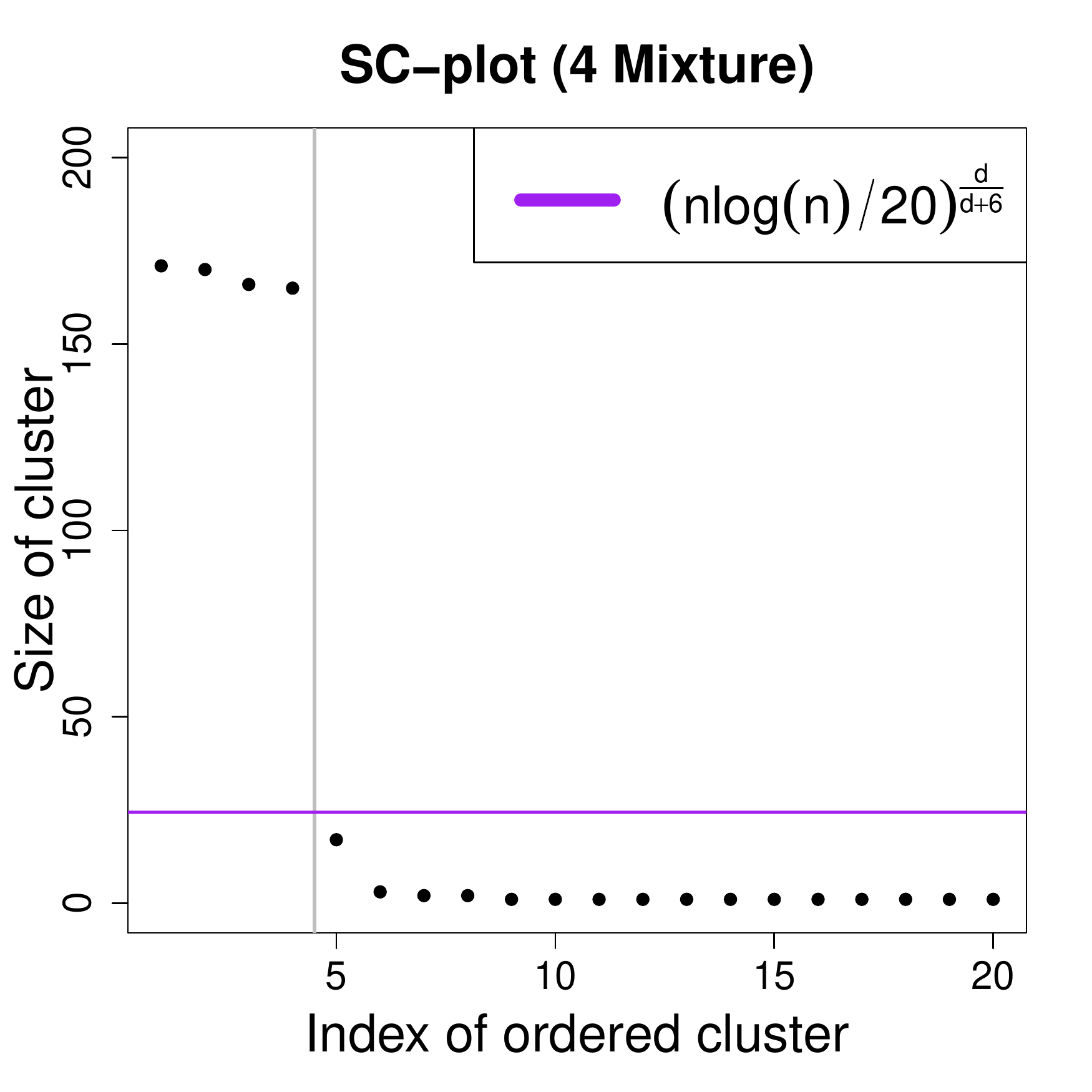} 
	}
\caption{An example of cluster noise.
These data are from a 4-Gaussian mixture in $d=8$. 
Panel (a) shows the first two coordinates and we add 
Gaussian noise to other $6$ coordinates.
Panel (b) shows the ordered size of clusters
from mode clustering using Silverman's rule \eqref{eq::h0}.
On the left side of gray line in panel (b) are the real clusters;
on the right side of gray line are the clusters we want to filter out.}
\label{Fig::Filter1}
\end{figure}

Figure~\ref{Fig::Filter1} gives an example on the small clusters
from a 4-Gaussian mixture and each mixture component contains $200$
points.
Note that this mixture is in $d=8$ and 
the first two coordinates are given in panel (a) of Figure~\ref{Fig::Filter1}.
Panel (b) shows the ordered size of clusters when the smoothing
parameter $h$ is chosen by the Silverman's rule (SR) given in \eqref{eq::h0}.
On the left side of the gray vertical line, the four clusters
are real signals while the clusters on the right hand side of the gray line 
are small clusters that we want to filter out.

\begin{figure}
\centering
\includegraphics[width=2 in, height=2 in]{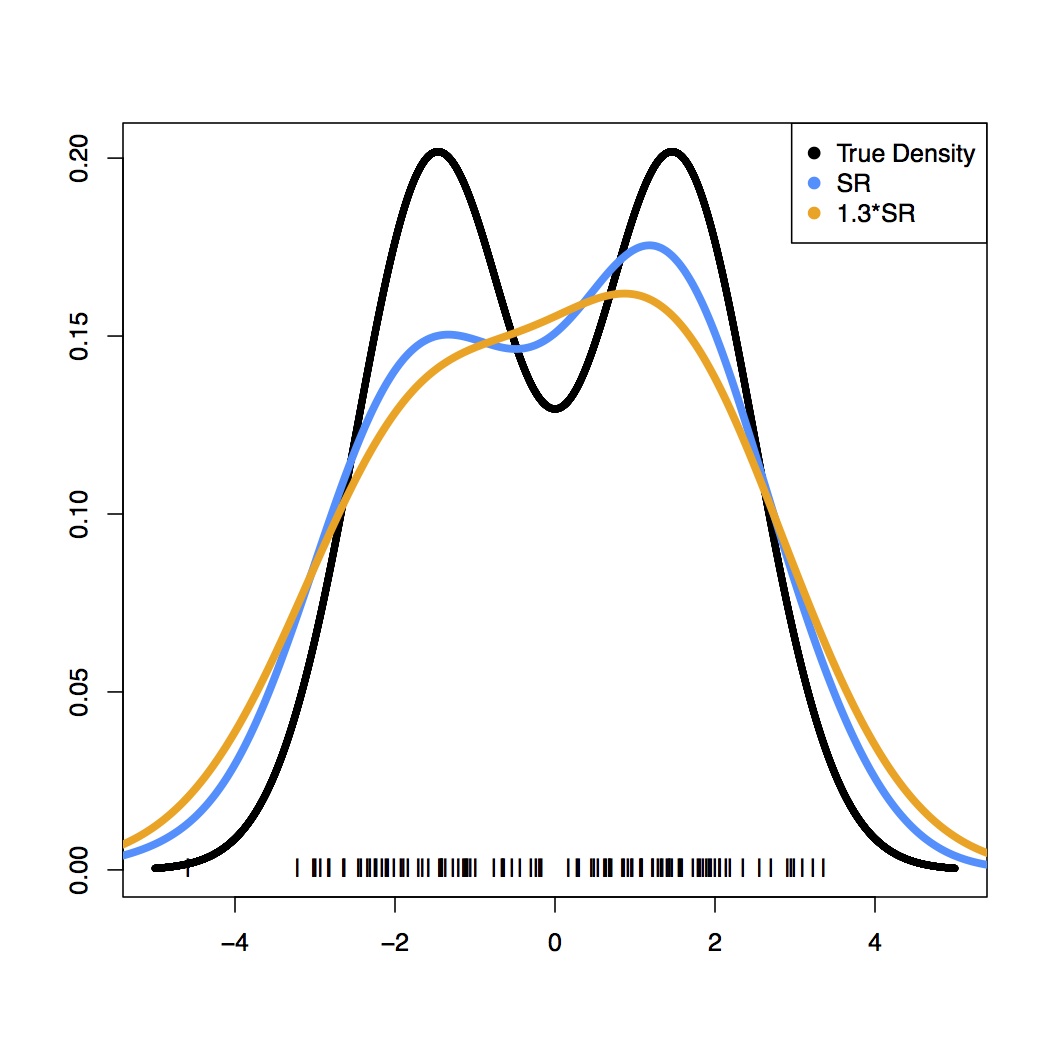}
\caption{An example for showing the problem of oversmoothing. 
This is a $n=100$ sample from a simple two Gaussian mixture in $d=1$. 
The black curve is the true density, the blue curve is the estimated
density based on the Silverman's rule (denoted as SR; see \eqref{eq::h0}) 
and the orange curve is $h= 1.3\times$ \eqref{eq::h0}. 
If we oversmooth too much (orange curve), we only
identify one cluster (mode).
}
\label{Fig::filter2}
\end{figure}

There are two approaches to deal with the clustering noise: 
increasing the smoothing parameters 
and merging (or eliminating) small clusters.
However, increasing the bandwidth oversmooths
which may wash out useful information. 
See Figure~\ref{Fig::filter2} for an example.
Thus, we focus on the method of
merging small clusters.
Our goal is to have a quick, simple method.

A simple merging method is to enforce a threshold  $n_0$
on the cluster size (i.e., number of data points within)
and merge points within the small clusters (size less than $n_0$)
into some nearby clusters whose size is larger or equal to $n_0$.
We will discuss how to merge tiny clusters latter.
Clusters with size larger or equal to $n_0$ are called
``significant'' clusters and those with size less than $n_0$ are called ``insignificant''
clusters. 
We recommend setting
\begin{equation}
n_0 = \left(\frac{n\log (n)}{20 }\right)^{\frac{d}{d+6}}.
\label{eq::n0}
\end{equation}
The intuition for the above rule is from the optimal $L_\infty$ error rate
for estimating the gradient (recall \eqref{eq::sup}).
The constant $20$ in the denominator is based on our experience
from simulations and later,
we will see that this rule works quiet well in practice.

\begin{figure}
\centering
	\subfigure[SC-plot for 4-Gaussian in 8-D]
	{
		\includegraphics[width=2 in, height=2 in]{figures/4Gmix_SC}
	}
	\subfigure[SC-plot for 5-clusters in 10-D]
	{
		\includegraphics[width=2 in, height=2 in]{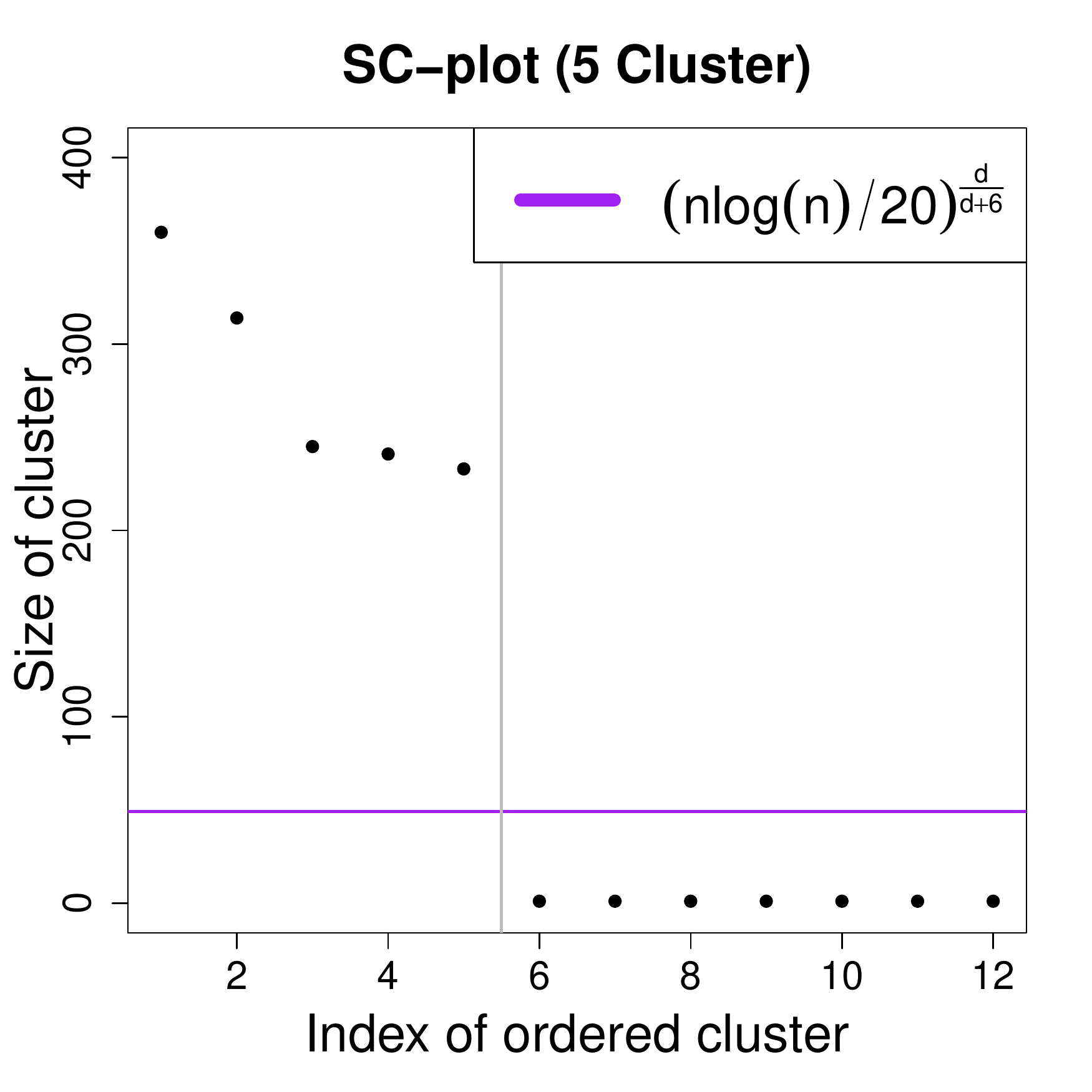} 
	}
\caption{The SC-plot for the 4-Gaussian example and the 5-clusters in 10-D example.
Notice that there is always a gap on the size of clusters near the gray line 
(boundary of real clusters and clustering noise).
This gap can be used to select the filtering threshold $n_0$.
}
\label{Fig::Filter3}
\end{figure}

Here we introduce the \emph{SC-plot} (Size of Cluster plot)
as a diagnostic for the choice of $n_0$.
The SC-plot displays the ordered size of clusters.
Ideally, there will be a gap
between the size of significant clusters
and insignificant clusters which in turns induces an elbow in the SC-plot.
Figure~\ref{Fig::Filter3} show the SC-plot
for a 4-Gaussian mixture in 8-dimension (the data used in Figure~\ref{Fig::Filter1}) 
and a 5-clusters in 10-dimension data (see section \ref{sec::5C} for more details).
Both data sets are simulated so that we know the true number of clusters 
(we use the gray line to separate clustering noise and real clusters).
Our reference rule \eqref{eq::n0} successfully separates the noise and signals
in both cases.
Note that SC-plot itself provides a summary of the structure of clusters.

After identifying tiny clusters, we use the following procedure to 
merge points within small clusters
(suggested to us by Jose Chacon).
We first remove points in tiny clusters and then use the remaining 
data (we call this the ``reduced dataset'') to estimate the density and 
perform mode clustering.
Since the reduced dataset does not include points within tiny clusters,
in most cases, this method outputs only stable clusters.
If there are still tiny clusters after merging,
we identify those points within tiny clusters 
and merge them again to other large clusters.
We repeat this process until there are no tiny clusters.
By doing so, we will cluster all data points into significant clusters.

\vspace{1cm}

{\bf Remark.}
In addition to the denoising method proposed above,
we can remove the clustering noise using persistent homology
\citep{chazal2011persistence}.
The threshold level for persistence can be computed
via the bootstrap \citep{Fasy2014}.
However, we found that this did not work well
except in low dimensions.
Also, it is extremely computationally intensive.

\section{Visualization}	\label{sec::V}

Here we present a method 
for visualizing the clusters
that combines multidimensional
scaling (MDS) with our connectivity measure for clusters.

\subsection{Review of Multidimensional Scaling}

Given points $X_1,\ldots,X_n\in\R^d$, 
classical MDS finds
$Z_1,\ldots,Z_n\in\R^k$ such that 
they minimize
\begin{equation}
\sum_{i,j}\left| (Z_i-\bar{Z}_n)^T(Z_j-\bar{Z}_n)-(X_i-\bar{X}_n)^T(X_j-\bar{X}_n) \right|^2.
\end{equation}

Note $\bar{Z}_n=\frac{1}{n}\sum_{i=1}^nZ_i$.  A nice feature for
classical MDS is the existence of a closed-form solution to $Z_i$'s.
Let $\mathbf{S}$ be a $n\times n$ matrix with
element $$\mathbf{S}_{ij} = (X_i-\bar{X}_n)^T(X_j-\bar{X}_n).$$ Let
$\lambda_1>\lambda_2>\cdots>\lambda_n$ be the eigenvalues of
$\mathbf{S}$ and $v_1,\ldots,v_n\in\R^n$ be the associated
eigenvectors. We denote $\mathbf{V}_k = [v_1,\ldots,v_k]$ and
$\mathbf{D}_k = \Diag(\sqrt{\lambda_1},\ldots,\sqrt{\lambda_k})$ be a
$k\times k$ diagonal matrix. Then it is known that each $Z_i$ is the
$i$-th row of $\mathbf{V}_k \mathbf{D}_k$
\citep{hastie01statisticallearning}. In our visualization, we
constrain $k=2$.


\subsection{Two-stage Multidimensional Scaling}

\begin{figure}
\centering
	\subfigure[MDS on modes]
	{
		\includegraphics[width=2 in, height=2 in]{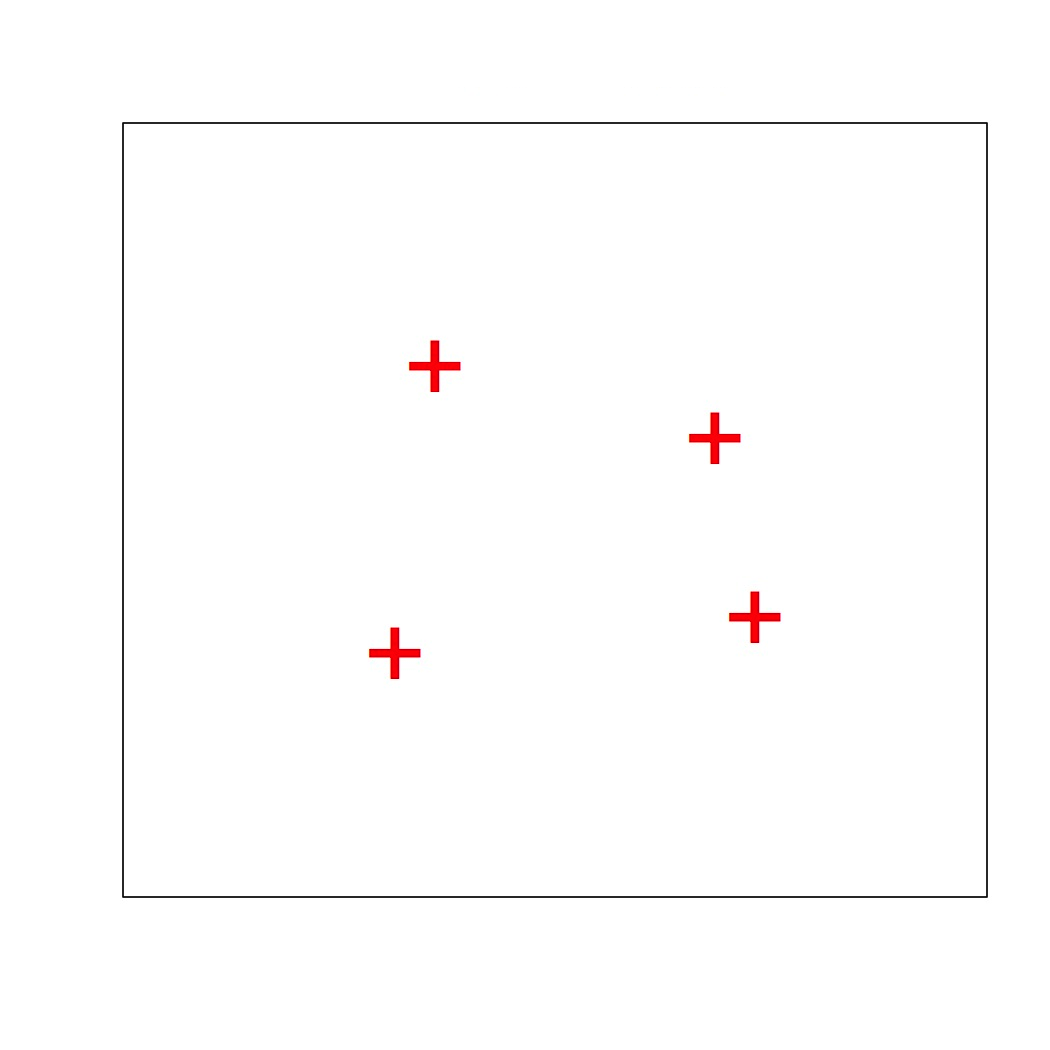}
	}
	\subfigure[MDS on each cluster]
	{
		\includegraphics[width=2 in, height=2 in]{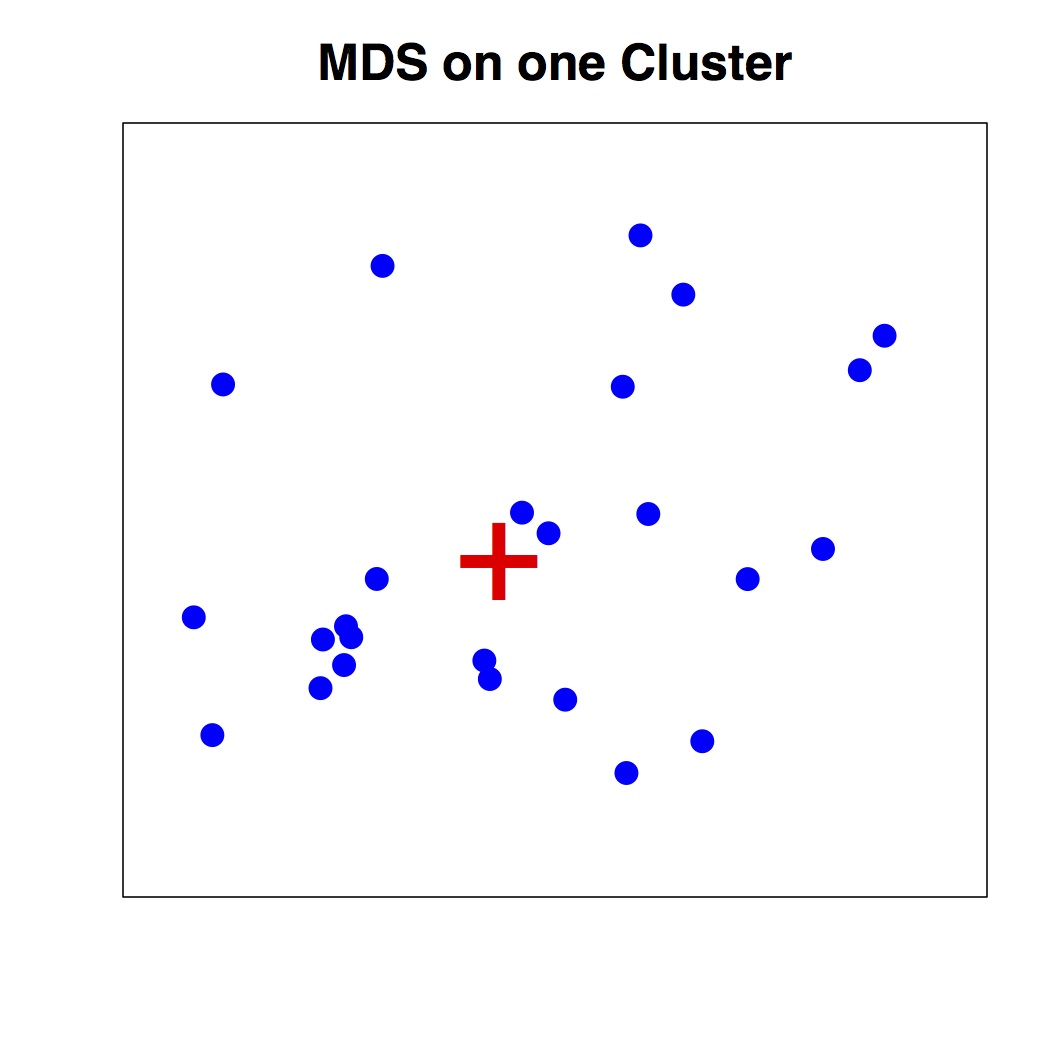} 
	}\\
	\subfigure[Place points around modes]
	{
		\includegraphics[width=2 in, height=2 in]{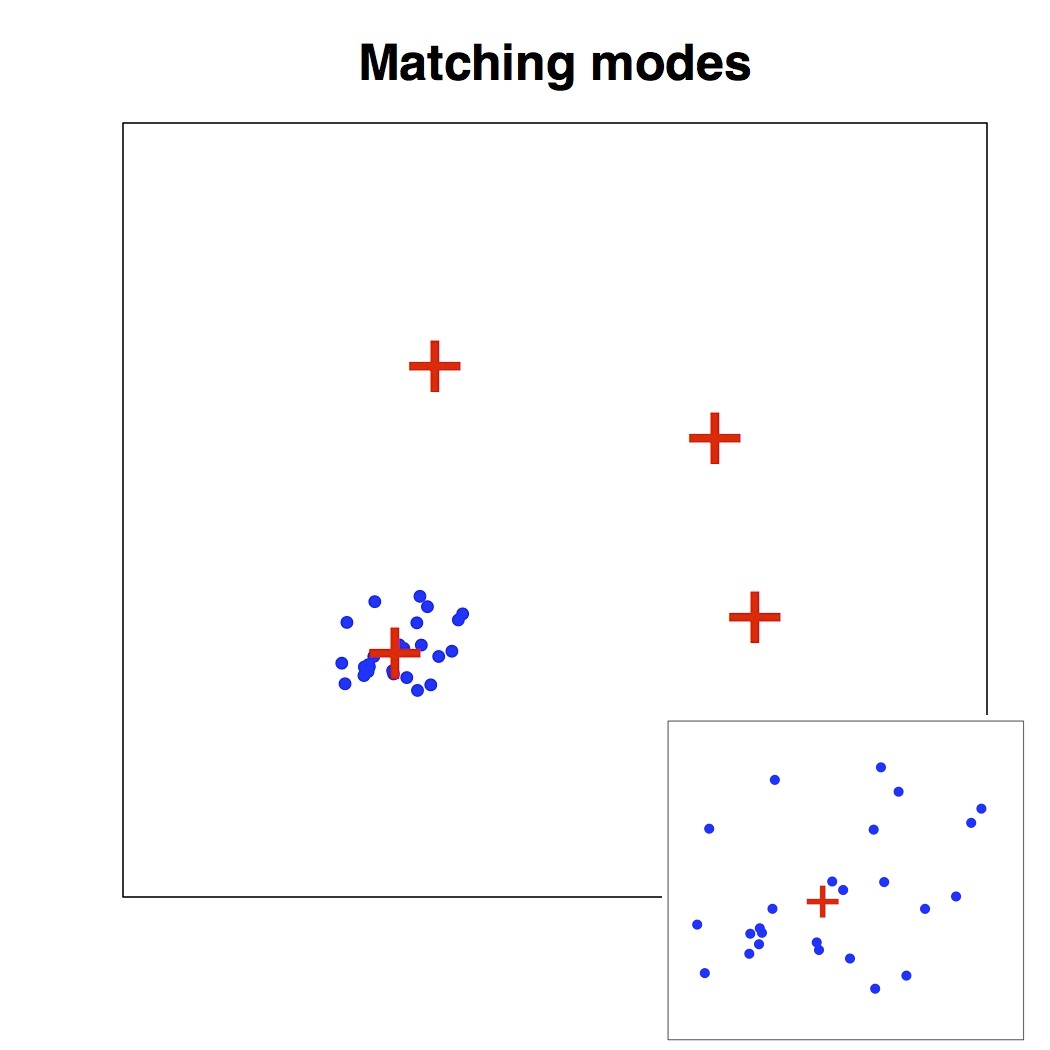}
	}
	\subfigure[The final result]
	{
		\includegraphics[width=2 in, height=2 in]{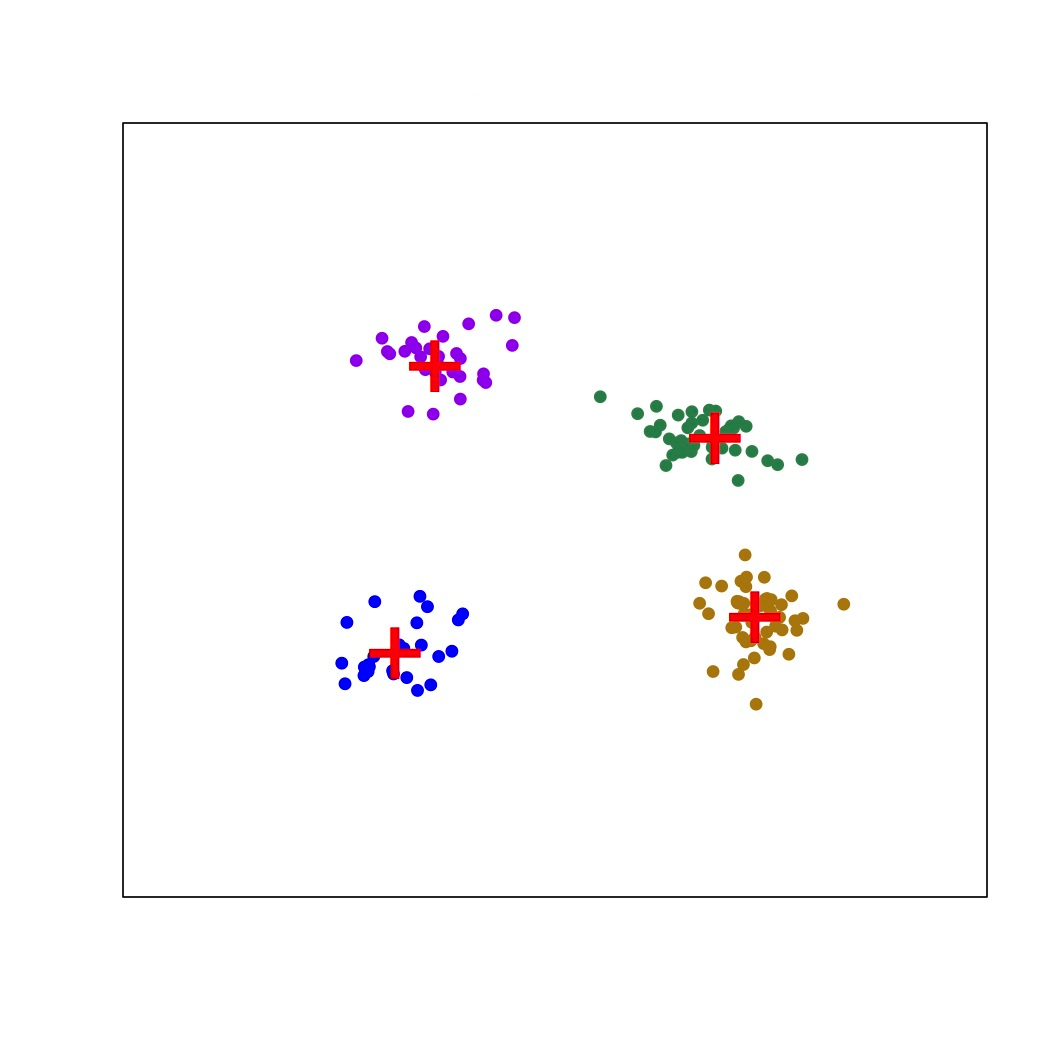} 
	}
\caption{An example for the two stage MDS. Note that the bottom right small plot in (c) is the plot in (b). At stage one, we run MDS for all modes and plot them as in (a). At stage two, we apply MDS for each cluster including the local mode as in (b). Then we place cluster points around local modes as in (c) and (d).}
\label{Fig::MDSEX}
\end{figure}

Our approach consists of two stages.
At the first stage, we apply MDS on the modes and plot the result in $\mathbb{R}^2$.
At the second stage, we apply MDS to points within each cluster 
along with the associated mode.
Then we place the points around the projected modes.
We scale the MDS result at the first stage by a factor $\rho_0$
so that each cluster is separated from each other. 
Figure~\ref{Fig::MDSEX} gives an example.

Recall that
$\hat\cM = \{\hat{m}_1,\ldots,\hat{m}_{\hat k}\}$ is the set of
estimated local modes and $\mathcal{X}_j$ is the set of data points
belonging to mode $\hat{m}_j$.  At the first stage, we perform MDS on
$\hat\cM$ so that
\begin{equation}
\{\hat{m}_1,\ldots,\hat{m}_{\hat{k}}\}\overset{\MDS}{\Longrightarrow}
\{\hat{m}^{\dagger}_1,\ldots,\hat{m}^{\dagger}_{\hat{k}}\},
\end{equation}
where $\hat{m}^{\dagger}_j\in \R^2$ for $j=1,\ldots,\hat{k}.$ We plot
$\{\hat{m}^{\dagger}_1,\ldots,\hat{m}^{\dagger}_{\hat{k}}\}$.

At the second stage, we consider each cluster individually. Assume we
are working on the $j$-th cluster and $\hat{m}_j,\mathcal{X}_j$ are the
corresponding local mode and cluster points. We denote $\mathcal{X}_j =
\{X_{j1},\ldots,X_{jN_j}\}$, where $N_j$ is the sample size for
cluster $j$. Then we apply MDS to the collection of points $\{m_j,
X_{j1}, X_{j2},\ldots,X_{jN_j}\}$:
\begin{equation}
\{m_j, X_{j1}, X_{j2},\ldots,X_{jN_j}\}\overset{\MDS}{\Longrightarrow}\{m^*_j, X^*_{j1}, X^*_{j2},\ldots,X^*_{jN_j}\},
\end{equation}
where $m^*_j, X^*_{j1}, X^*_{j2},\ldots,X^*_{jN_j}\in\R^2$.  Then we
center the points at $\hat{m}^{\dagger}_j$ and place $X^*_{j1},
X^*_{j2},\ldots,X^*_{jN_j}$ around $\hat{m}^{\dagger}_j$. That is, we
make a translation to the set $\{m^*_j, X^*_{j1},
X^*_{j2},\ldots,X^*_{jN_j}\}$ so that $m^*_j$ matches the location of
$\hat{m}^{\dagger}_j$. Then we plot the translated points $X^*_{j1},
X^*_{j2},\ldots,X^*_{jN_j}$. We repeat the above process for each
cluster to visualize the high dimensional clustering.

Note that in practice, the above process may cause unwanted overlap among
clusters.  
Thus, one can scale
$\{\hat{m}^{\dagger}_1,\ldots,\hat{m}^{\dagger}_{\hat{k}}\}$ by a
factor $\rho_0>1$ to remove the overlap.

One can use other dimension reduction techniques as well.
For instance,
we can use the
landmark MDS \citep{silva2002global, de2004sparse}
and treats each local mode as the landmark points.
This provides an alternative way to visualize the clusters.

\subsection{Connectivity Graph}

We can improve the visualization of the previous subsection
by accounting for the connectivity of the clusters.
We apply the connectivity measure introduced in section \ref{sec::connect}.
Let $\hat{\Omega}$ be the matrix for the connectivity measure defined in \eqref{eq::CC2}.
We connect two clusters, say $i$ and $j$, by a straight line if the connectivity measure $\hat{\Omega}_{ij}>\omega_0$,
a pre-specified threshold. Our experiments show that 
\begin{equation}
\omega_0 =\frac{1}{2\times\mbox{number of clusters}}
\label{eq::omega0}
\end{equation} 
is a good default choice. We can adjust the width of the connection line 
between clusters to show the strength of connectivity.
See Figure~\ref{Fig::VH2} panel (a) for an example; the edge
linking clusters (2,3) is much thicker than any other edge.

Algorithm \ref{alg::Visualization} summarizes the process of visualizing
high dimensional clustering. Note that the smoothing bandwidth $h$ and
the thresholding of cluster size $n_0$ can be chosen by the methods
proposed in section~\ref{sec::h}. The remaining two parameters $\rho_0$ and
$\omega_0$ are visualization parameters; they are not involved in any
analysis so that one can change these parameters freely.

\begin{algorithm}
\caption{Visualization for Mode Clustering}
\begin{algorithmic}
\State \textbf{Input:} Data $\mathbb{X} =\{X_i:  i=1,\ldots,n\}$, 
bandwidth $h$, parameters $n_0$, $\rho_0$, $\omega_0$.
\vspace{0.1 in}
\State \textbf{Phase 1:} Mode clustering
\State 1. Use the mean shift algorithm for clustering based on bandwidth $h$.
\State 2. (Optional) Find clusters of size less than $n_0$ and merge them with larger clusters.

\vspace{0.1 in}
\State \textbf{Phase 2:} Dimension reduction
\State Let $\{(m_j,\mathcal{X}_j): j=1,\ldots,\hat{k}\}$ be the pairs of local modes and the associated data points.
\State 3. Perform MDS to each $(\hat{m}_j,\mathcal{X}_j)$ to get $(\hat{m}^*_j,\mathcal{X}^*_j)$.
\State 4. Perform MDS to modes only to get $\{\hat{m}^{\dagger}_1,\ldots,\hat{m}^{\dagger}_{\hat{k}}\}$.
\State 5. Place each $\hat{M}^{\dagger}_j=\rho_0 \times \hat{m}^{\dagger}_j$ on the reduced coordinate.
\State 6. Place each $(\hat{m}_j,\mathcal{X}_j)$ around $\hat{M}^{\dagger}_j$ by matching $\hat{m}_j\rightarrow \hat{M}^{\dagger}_j$.

\vspace{0.1 in}
\State \textbf{Phase 3:} Connectivity measure
\State 7. Estimate $\Omega_{ij}$ by \eqref{eq::CC2} and one of the above soft clustering methods.
\State 8. Connect $\hat{M}^{\dagger}_i,\hat{M}^{\dagger}_j$ if $\hat{\Omega}_{ij}>\omega_0$.
\end{algorithmic}
\label{alg::Visualization}
\end{algorithm}

\section{Experiments}	\label{sec::ex}

\begin{figure}
\centering
	\includegraphics[scale=0.4]{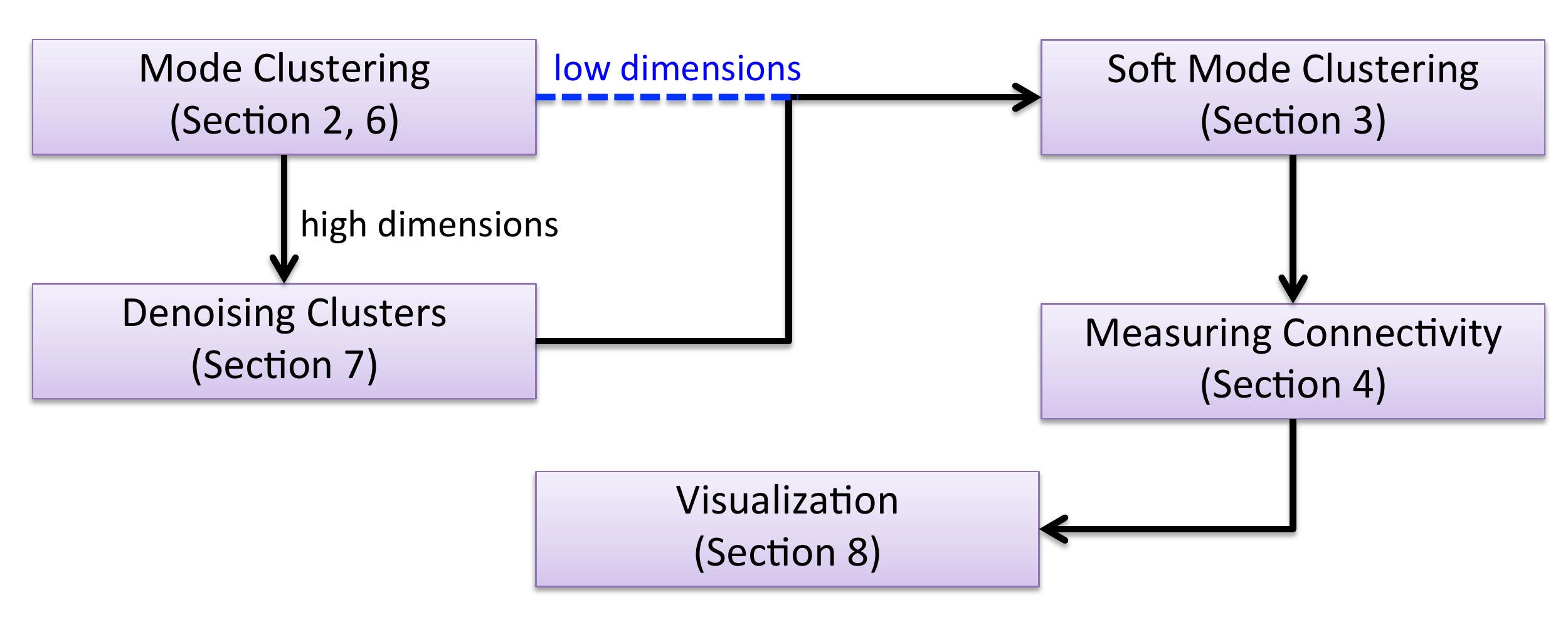}
\caption{A flowchart for the clustering analysis using proposed methods.
This shows a procedure to conduct a high-dimensional clustering
using the proposed methods in the current papers.
We apply this procedure to the data in section \ref{sec::5C} to \ref{sec::seeds}.
}
\label{Fig::flowchart}
\end{figure}

We present several experiments in this section. The 
parameters were chosen as follows:
we choose $h$ based on
\eqref{eq::h0}, $\omega_0$ based on
\eqref{eq::omega0}.
Figure~\ref{Fig::flowchart} gives a flowchart that summarizes clustering 
analysis using the approach presented in this paper.
Given the multivariate data, we first select the bandwidth
and then conduct (hard) mode clustering.
Having identified clusters, we denoise small clusters by merging them into significant clusters
and apply soft-mode clustering to measure the connectivity.
Finally, we visualize the data using the two-step MDS approach
and connect clusters if the pairwise connectivity is high.
This establishes a procedure for multivariate clustering
and we apply it to the data in
Sections \ref{sec::5C} to \ref{sec::seeds}.

\subsection{5-clusters in 10-D}	\label{sec::5C}

\begin{figure}
\centering
	\subfigure[The first three coordinates.]
	{
		\includegraphics[width=2 in, height=2 in]{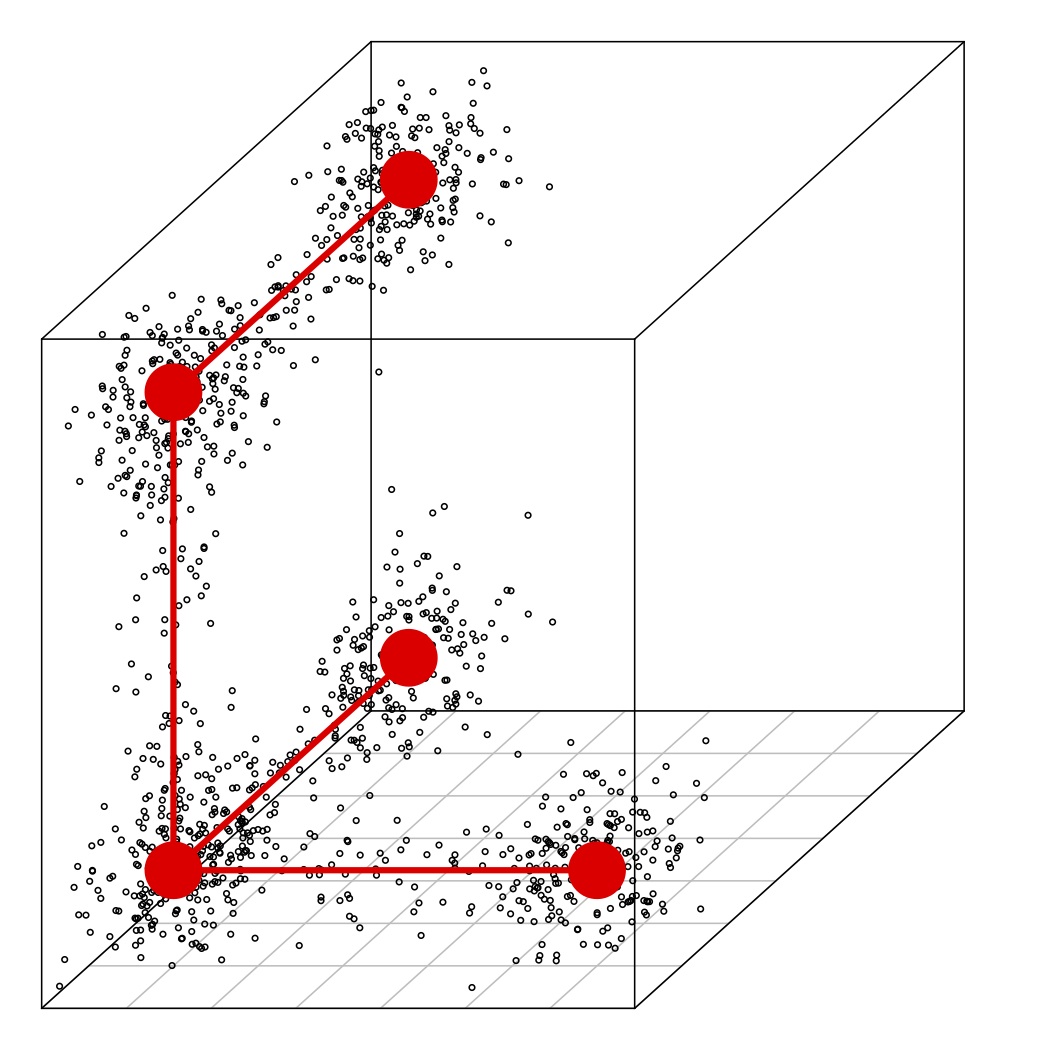}
	}

	\subfigure[Visualization]
	{
		\includegraphics[width=2 in, height=2 in]{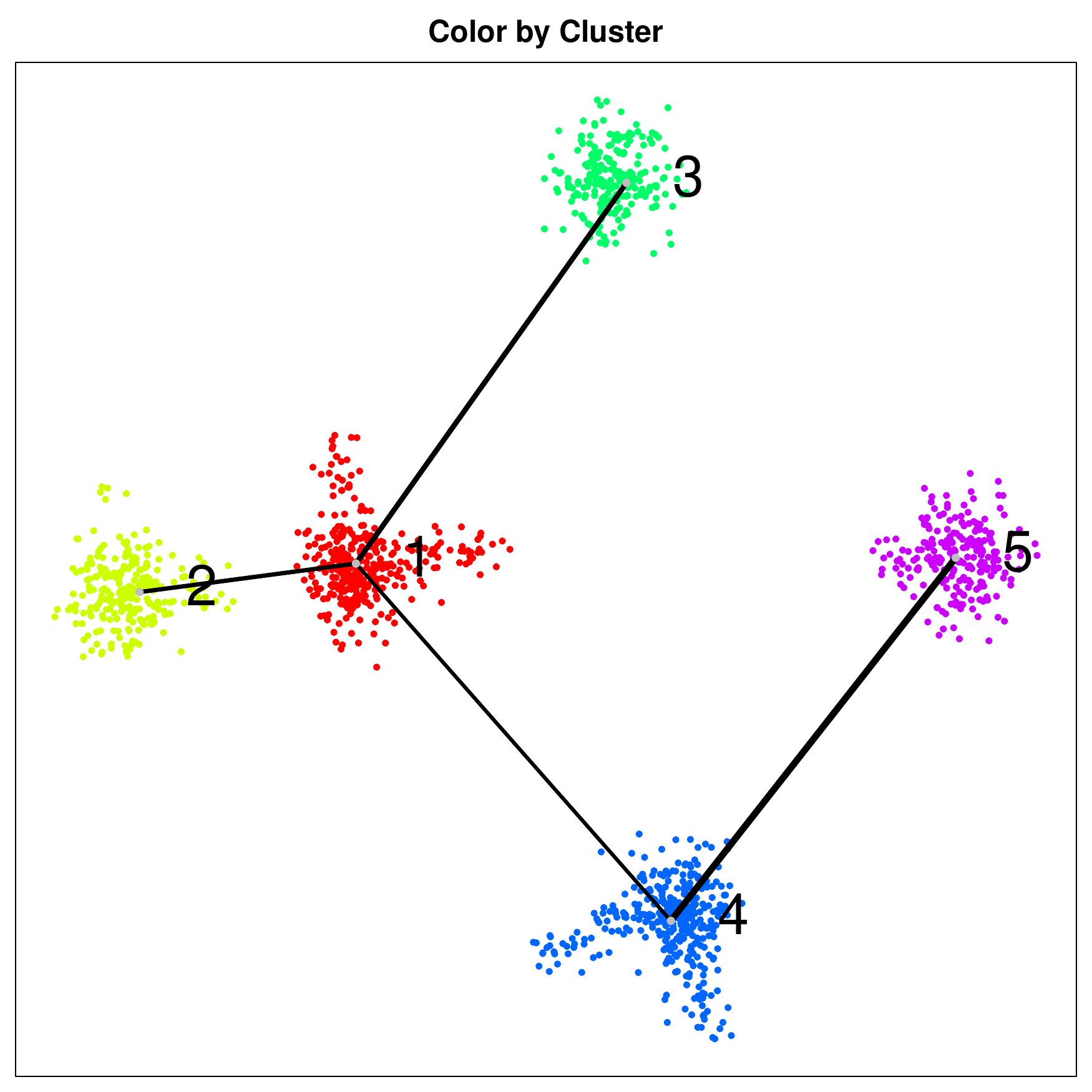} 
	}\\
	\subfigure[Matrix of connectivity]
	{
\begin{tabular}{rrrrrr}
  \hline
 & 1 & 2 & 3 & 4 & 5 \\ 
  \hline
1 & -- &\color{red} 0.15 &\color{red} 0.14 &\color{red} 0.12 & 0.02 \\ 
  2 &\color{red} 0.15 & -- & 0.03 & 0.03 & 0.00 \\ 
  3 &\color{red} 0.14 & 0.03 & -- & 0.02 & 0.00 \\ 
  4 &\color{red} 0.12 & 0.03 & 0.02 & -- &\color{red} 0.16 \\ 
  5 & 0.02 & 0.00 & 0.00 &\color{red} 0.16 &-- \\ 
   \hline
\end{tabular}
	}
\caption{Visualization of clustering on the 10-dimensional 5-cluster data. This is a 5 cluster data with `filament' connecting them in $d=10$. Panel (a) shows the first three coordinates
(which contains real structures; the rest $7$ dimensions are Gaussian noise).}
\label{Fig::VH}
\end{figure}

We implement our visualization technique in the following `5-cluster' data.
We consider $d=10$ and $5$ Gaussian mixture centered at the following positions
\begin{equation}
\begin{aligned}
C_1 &= (0,0,0,0,0,0,0,0,0,0)\\
C_2 &= (0.1,0,0,0,0,0,0,0,0,0)\\
C_3 &= (0,0.1,0,0,0,0,0,0,0,0)\\
C_4 &= (0,0,0,0.1,0,0,0,0,0,0)\\
C_5 &= (0,0.1,0.1,0,0,0,0,0,0,0).
\end{aligned}
\end{equation}
For each Gaussian component, we generate $200$ data points from
$\sigma_1= 0.01$ and each Gaussian is isotropically distributed.
Then we consider four ``edges'' connecting pairs of centers. These edges
are $E_{12},E_{13},E_{14},E_{45}$, where $E_{ij}$ is the edge between
$C_i,C_j$.  We generate $100$ points from an uniform distribution over
each edge and add an isotropic iid Gaussian noise to each edge with
$\sigma_2 = 0.005$.  Thus, the total sample size is $1,400$ and
consist of 5 clusters centered at each $C_i$ and part of the clusters
are connected by a `noisy path' (also called filament
\citep{Genovese2012a, Chen2014}). 
The density has structure only at
the first three coordinates; a visualization for the structure 
is given in Figure \ref{Fig::VH}-(a).

The goal is to identify the five clusters as well as their
connectivity.  We display the visualization and the connectivity
measures in Figure \ref{Fig::VH}. 
All the parameters used in this
analysis is given as follows.
$$
h= 0.0114,\quad n_0= 49.05,\quad \quad\rho_0=2, \quad \omega_0=0.1.
$$
Note that the filtering threshold $n_0$ is picked by \eqref{eq::n0} and the SC-plot is
given in Figure \ref{Fig::Filter3} panel (b).


\subsection{Olive Oil Data}	\label{sec::OO}

We apply our methods to the Olive Oil data introduced in
\cite{Forina1983}. This data set consists of 8 chemical measurements (features)
for each observation and the total sample size is $n=572$. Each
observation is an olive oil produced in one of 3 regions in Italy and these
regions are further divided into 9 areas. Some other analyses
for this data can be found in \cite{Stuetzle:2003hb,Azzalini:2007la}. We hold out the information of
the areas and regions and use only the 8 chemical measurement to
cluster all the data.

\begin{figure}
\centering
\includegraphics[width=2 in, height=2 in]{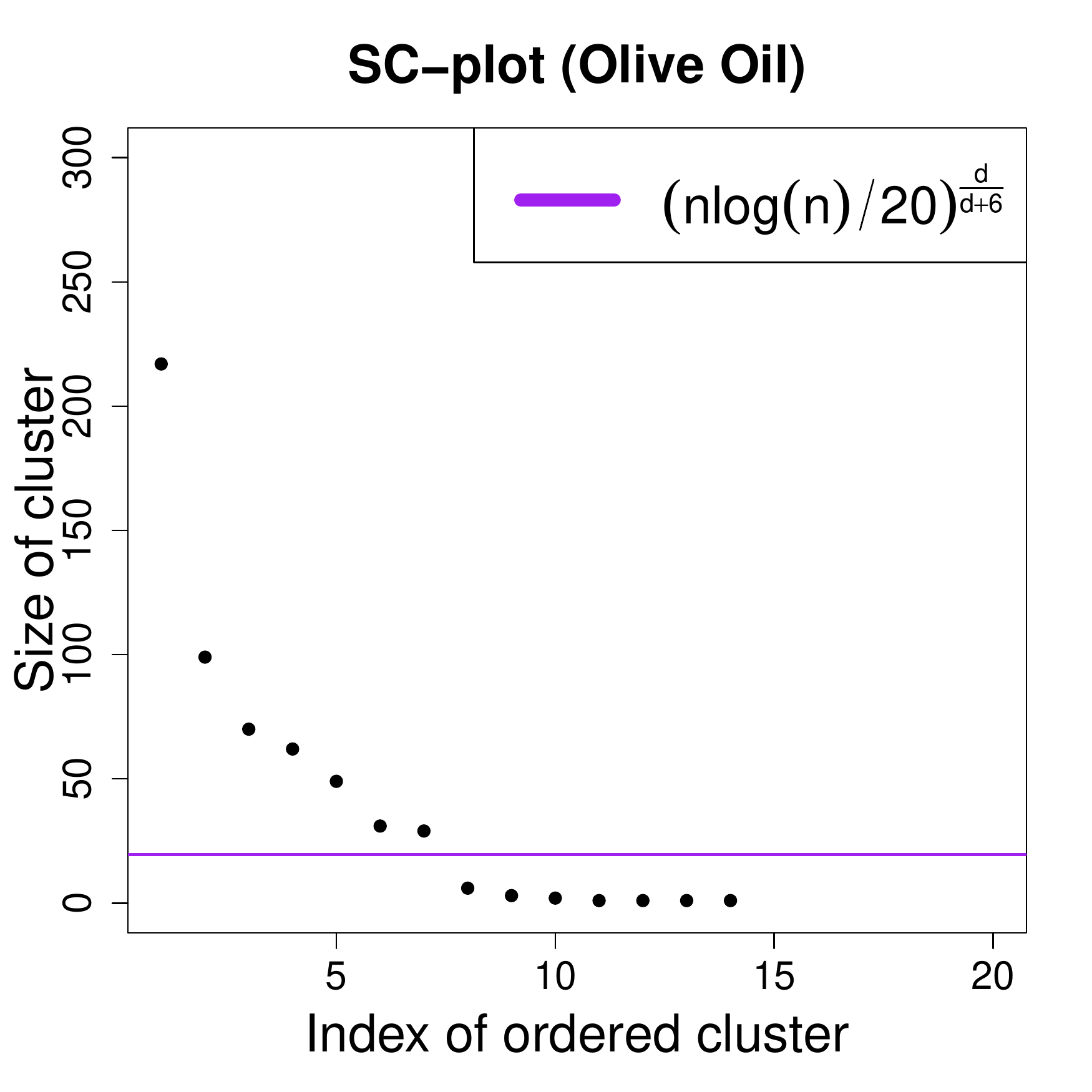} 
\caption{The SC-plot for Olive Oil data. 
The threshold $n_0= 19.54$ is within the a gap ($29$ to $6$) 
between size of seventh and eighth cluster.}
\label{Fig::OliveSC}
\end{figure}

\begin{figure}
\centering
	\subfigure[Visualization]
	{
		\includegraphics[width=2 in, height=2 in]{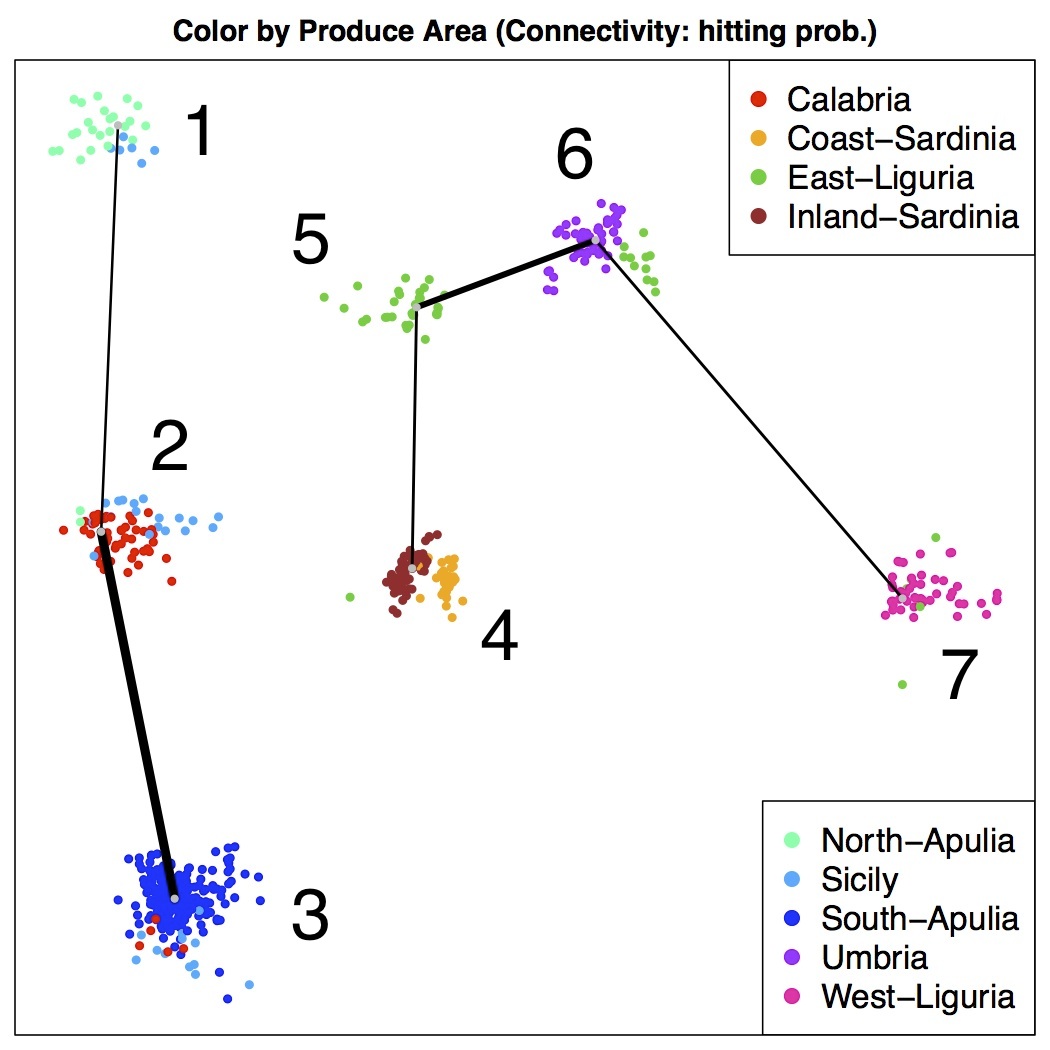} 
	}\\
	\subfigure[Map of produce area]
	{
	\includegraphics[height=2.5 in]{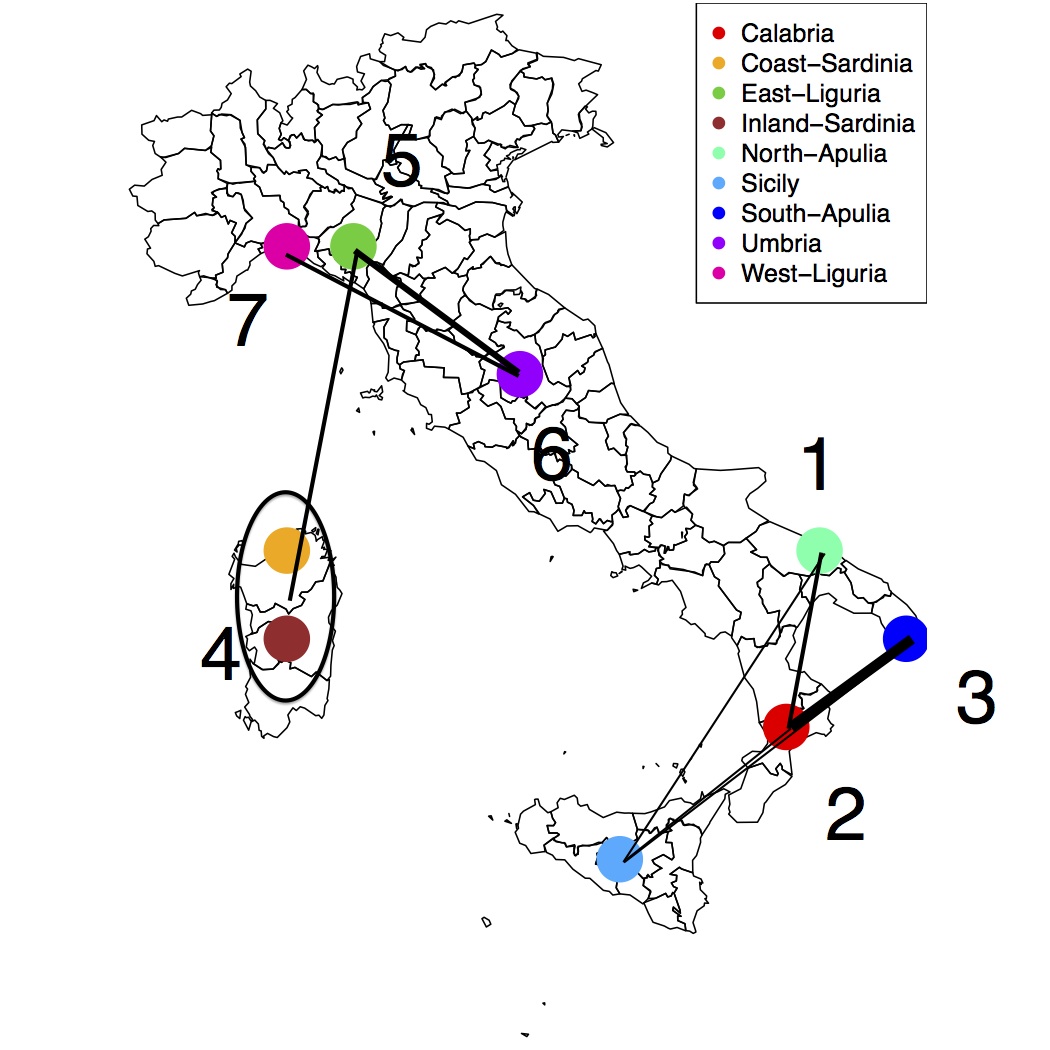} 
	}

\caption{(a): Clustering result for the Olive oil data ($d=8$). 
Note that we add edges to those pairs of clusters with
connectivity measure $>0.07$ (colored by red in the matrix). 
The connectivity matrix is in Figure~\ref{Fig::VH2_1}.
The width of the edge reflects the degree of connection. 
(b): The corresponding map of Italy. 
We assign the cluster label to the dominating
produce area and connect the edge according to the connectivity matrix. 
Note that the Sicily is spread out over cluster 1-3 so that we 
use dash lines to connect Sicily to Calabria, South-Apulia and North-Apulia.}
\label{Fig::VH2}
\end{figure}

Since these measurements are in different units, we normalize and
standardize each measurement.  We apply \eqref{eq::h0} for selecting $h$ and
thresholding the size of clusters based on \eqref{eq::n0}. 
Figure~\ref{Fig::OliveSC}
shows the SC-plot and the gap occurs between the seventh (size: $29$) 
and eighth cluster (size: $6$) and our threshold $n_0=19.54$ is within this gap.
We move the insignificant clusters into the nearest significant clusters.
After filtering, 7 clusters remain and 
we apply algorithm~\ref{alg::Visualization} for visualizing these clusters.  
To measure the connectivity, we apply the hitting probability so that we
do not have to choose the constant $\beta_0$. 
To conclude, we use the following
parameters
$$
h= 0.587, \quad n_0=19.54, \quad \rho_0=6,\quad\omega_0=0.071.
$$
The visualization is given in Figure~\ref{Fig::VH2}
and matrix of connectivity is given in Figure~\ref{Fig::VH2_1}. 
We color each point according to the produced area
to see how our methods capture the structure of data.

As can be seen, most clusters contain one dominating type of olive oil
(type: produce area).  Even in cases where one cluster contains
multiple types of olive oil, the connectivity matrix captures this
phenomena.  For instance, cluster 2 and 3 both contain Calabria,
Sicily and South-Apulia.  We do observe a connection between cluster 2
and 3 in Figure~\ref{Fig::VH2_1} and a higher connectivity measure in
the matrix for connectivity measures.
We display the map of Italy in panel (b) of Figure \ref{Fig::VH2}.
Mode clustering and connectivity measures reflect
the relationship in terms of geographic distance.

As can be seen in Figure \ref{Fig::VH2}, the clustering indeed captures
the difference in produce area.  More importantly, the connectivity
measurement captures the hidden structures of the produce area in the
following sense.  When a group of oil produced in the same area is
separated into two clusters, we observe an edge between these two
clusters.  This shows that the connectivity measure conveys more
information on the hidden interaction between clusters.

\begin{figure}
\centering
	\subfigure[Produce area versus cluster.]
	{
	\begin{tabular}{rrrrrrrr}
		  \hline
		 & 1 & 2 & 3 & 4 & 5 & 6 & 7 \\ 
		  \hline
		Calabria & 0 & 51 & 5 & 0 & 0 & 0 & 0 \\ 
		  Coast-Sardinia & 0 & 0 & 0 & 33 & 0 & 0 & 0 \\ 
		  East-Liguria & 0 & 0 & 0 & 1 & 32 & 11 & 6 \\ 
		  Inland-Sardinia & 0 & 0 & 0 & 65 & 0 & 0 & 0 \\ 
		  North-Apulia & 23 & 2 & 0 & 0 & 0 & 0 & 0 \\ 
		  Sicily & 6 & 18 & 12 & 0 & 0 & 0 & 0 \\ 
		  South-Apulia & 0 & 0 & 206 & 0 & 0 & 0 & 0 \\ 
		  Umbria & 0 & 0 & 0 & 0 & 0 & 51 & 0 \\ 
		  West-Liguria & 0 & 0 & 0 & 0 & 0 & 0 & 50 \\ 
		   \hline
	\end{tabular}
	}\\
	\subfigure[Matrix of connectivity]
	{
	\begin{tabular}{rrrrrrrr}
		  \hline
		 & 1 & 2 & 3 & 4 & 5 & 6 & 7 \\ 
		  \hline
		1 & -- &\color{red} 0.08 & 0.05 & 0.00 & 0.01 & 0.02 & 0.00 \\ 
		  2 &\color{red} 0.08 & -- &\color{red} 0.30 & 0.01 & 0.01 & 0.00 & 0.00 \\ 
		  3 & 0.05 &\color{red} 0.30 & -- & 0.02 & 0.01 & 0.00 & 0.00 \\ 
		  4 & 0.00 & 0.01 & 0.02 & -- &\color{red} 0.09 & 0.02 & 0.01 \\ 
		  5 & 0.01 & 0.01 & 0.01 &\color{red} 0.09 & -- &\color{red} 0.19 & 0.04 \\ 
		  6 & 0.02 & 0.00 & 0.00 & 0.02 &\color{red} 0.19 & -- & \color{red}0.09 \\ 
		  7 & 0.00 & 0.00 & 0.00 & 0.01 & 0.04 & \color{red}0.09 & -- \\ 
		   \hline
	\end{tabular}
	}
\caption{Confusion matrix (produce area versus cluster) 
and matrix of connectivity for the Olive oil data ($d=8$). 
We mark edges with
connectivity measure $>0.07$ by red color.
}
\label{Fig::VH2_1}
\end{figure}

\subsection{Banknote Authentication Data}

\begin{figure}
\centering
	\subfigure[Visualization]
	{
		\includegraphics[width=2 in, height=2 in]{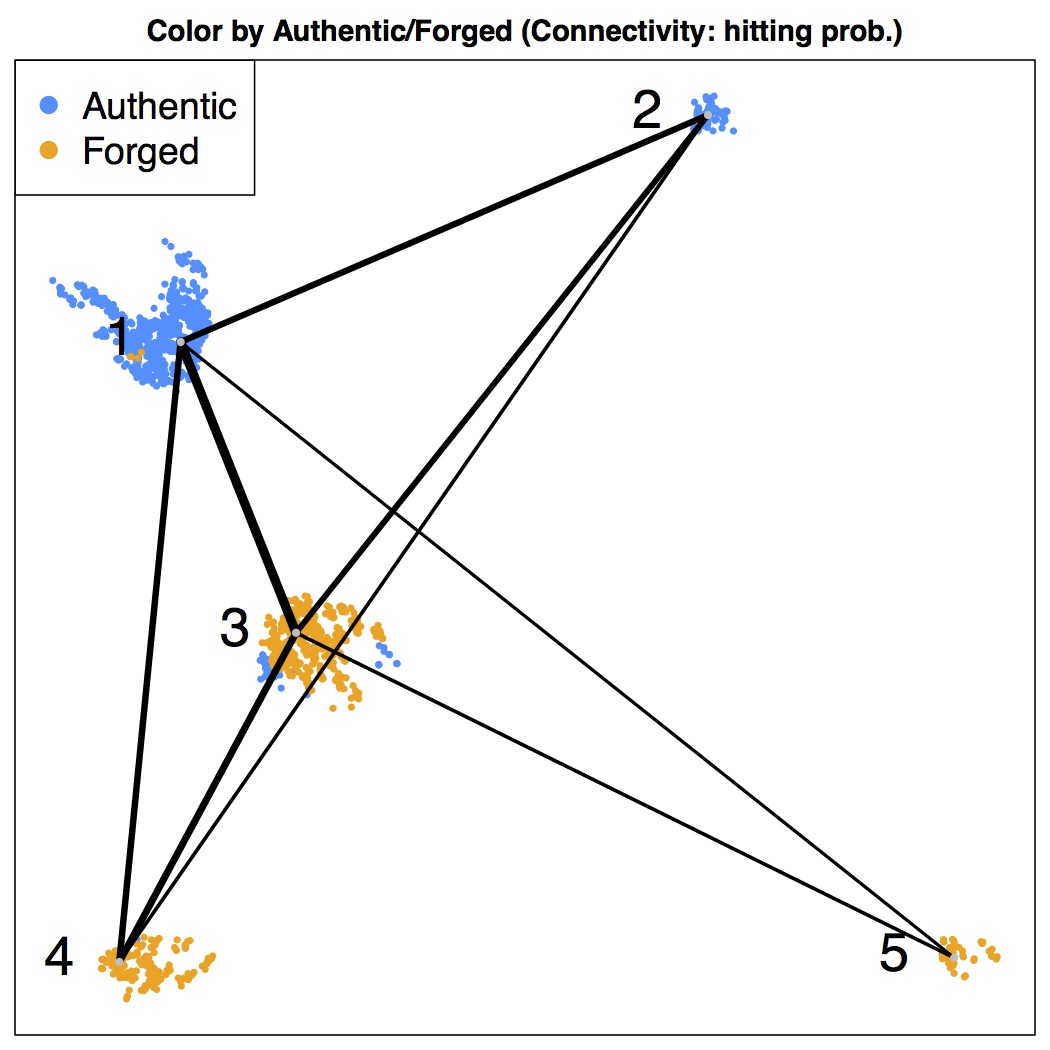} 
	}\\
	\subfigure[Authentic and Forged currency versus cluster.]
	{
\begin{tabular}{crrrrr}
  \hline
 & 1 & 2 & 3 & 4 & 5 \\ 
  \hline
Authentic & 629 &  70 &  62 &   1 &   0 \\ 
Forge &   4 &   0 & 390 & 179 &  37 \\ 
   \hline
\end{tabular}
	}\\
	\subfigure[Matrix of connectivity]
	{
\begin{tabular}{rrrrrr}
  \hline
 & 1 & 2 & 3 & 4 & 5 \\ 
  \hline
1 & -- &\color{red} 0.20 &\color{red} 0.30 &\color{red} 0.21 &\color{red} 0.11 \\ 
  2 &\color{red} 0.20 & -- &\color{red} 0.19 &\color{red} 0.12 & 0.06 \\ 
  3 &\color{red} 0.30 &\color{red} 0.19 & -- &\color{red} 0.22 &\color{red} 0.12 \\ 
  4 &\color{red} 0.21 &\color{red} 0.12 &\color{red} 0.22 & -- & 0.06 \\ 
  5 &\color{red} 0.11 & 0.06 &\color{red} 0.12 & 0.06 & -- \\ 
   \hline
\end{tabular}

	}
\caption{Clustering result for the Bank Authentication data ($d=4$). 
BlueViolet color is authentic banknote and orange color is the forged banknote.
The first two clusters are of the genuine classes while the latter three clusters
are the group of forged.
}
\label{Fig::Bank2}
\end{figure}

We apply our methods to the banknote authentication data set given in 
the UCI machine learning database repository \citep{asuncion2007uci}.
The data are extracted from images that are taken from authentic and forged banknote-like
specimens and later are digitalized via an industrial camera for print inspection.
Each image is a $400\times400$ pixels gray scale picture with a resolution of about 660 dpi.
A wavelet transform is applied to extract features from the images.
Each data point contains four attributes: `variance of Wavelet Transformed image',
`skewness of Wavelet Transformed image', `kurtosis of Wavelet Transformed image'
and `entropy of image'.

We apply our methods to analyze this dataset.
Note that all clusters are larger than $n_0=11.97$ 
(the smallest cluster has $37$ points)
so that we do not filter out any cluster.
The following
parameters are used:
$$
h= 0.613,\quad n_0=11.97, \quad \rho_0=5,\quad\omega_0=0.1.
$$
The visualization, confusion matrix and matrix of connectivity 
are given in Figure~\ref{Fig::Bank2}.

From Figure~\ref{Fig::Bank2}, cluster $1$ and $2$
are clusters for the real banknotes while cluster $3,4$ and $5$
are clusters of fake banknotes.
By examining the confusion matrix (panel (b)),
the cluster $2$ and $5$ are clusters for purely genuine and forged banknote.
As can be seen from panel (c), their connectivity is relatively small
compared to the other pairs.
This suggests that the authentic and fake banknotes
are really different in a sense.

\subsection{Wine Quality Data} 	\label{sec::WQ}

\begin{figure}
\centering
\includegraphics[width=2 in, height=2 in]{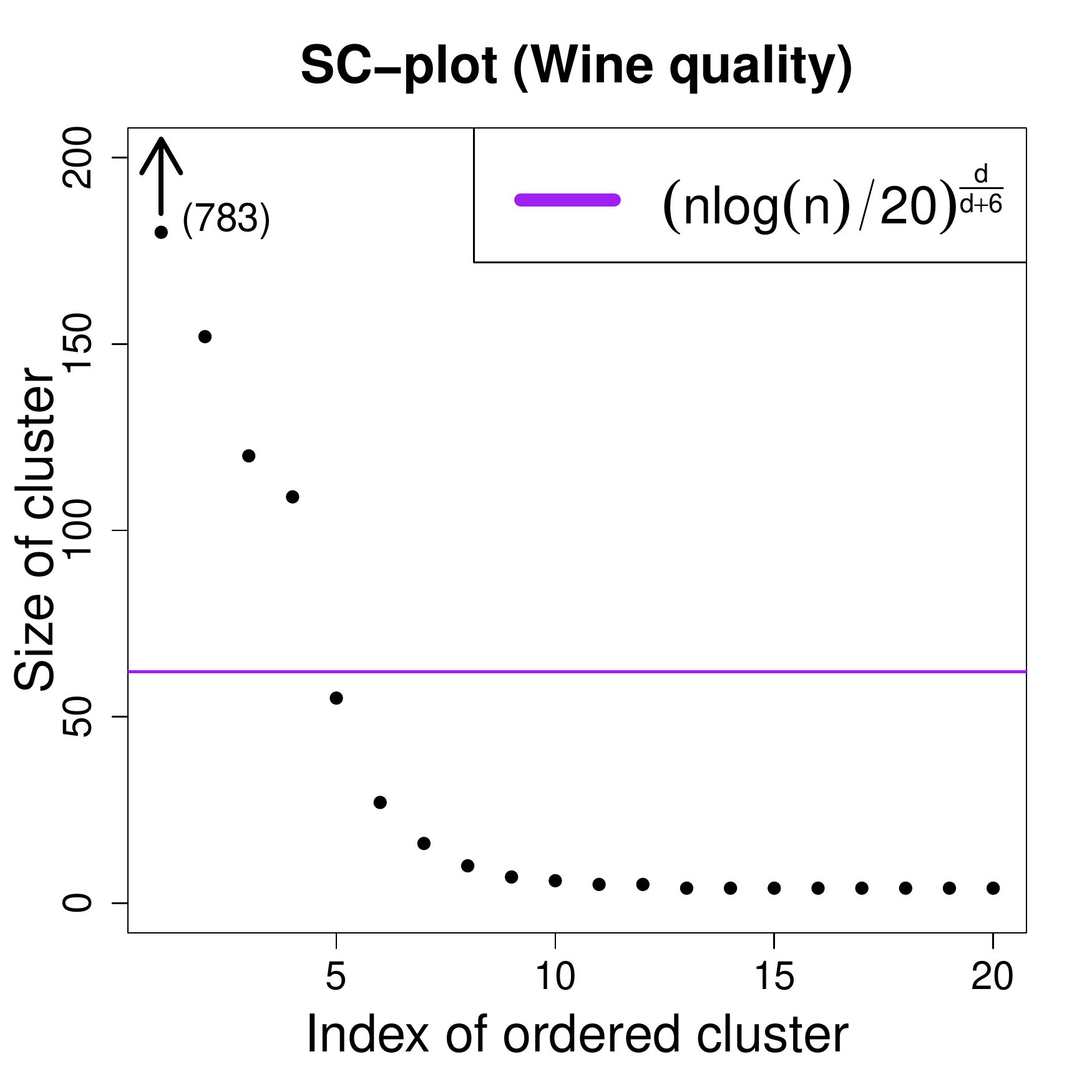} 
\caption{The SC-plot for wine quality data. 
Our choice of $n_0 = 62.06$ which agrees
with the gap between fourth and fifth cluster (containing $109$ and $55$ points).}
\label{Fig::WineSC}
\end{figure}

\begin{figure}
\centering
	\subfigure[Visualization]
	{
		\includegraphics[width=2 in]{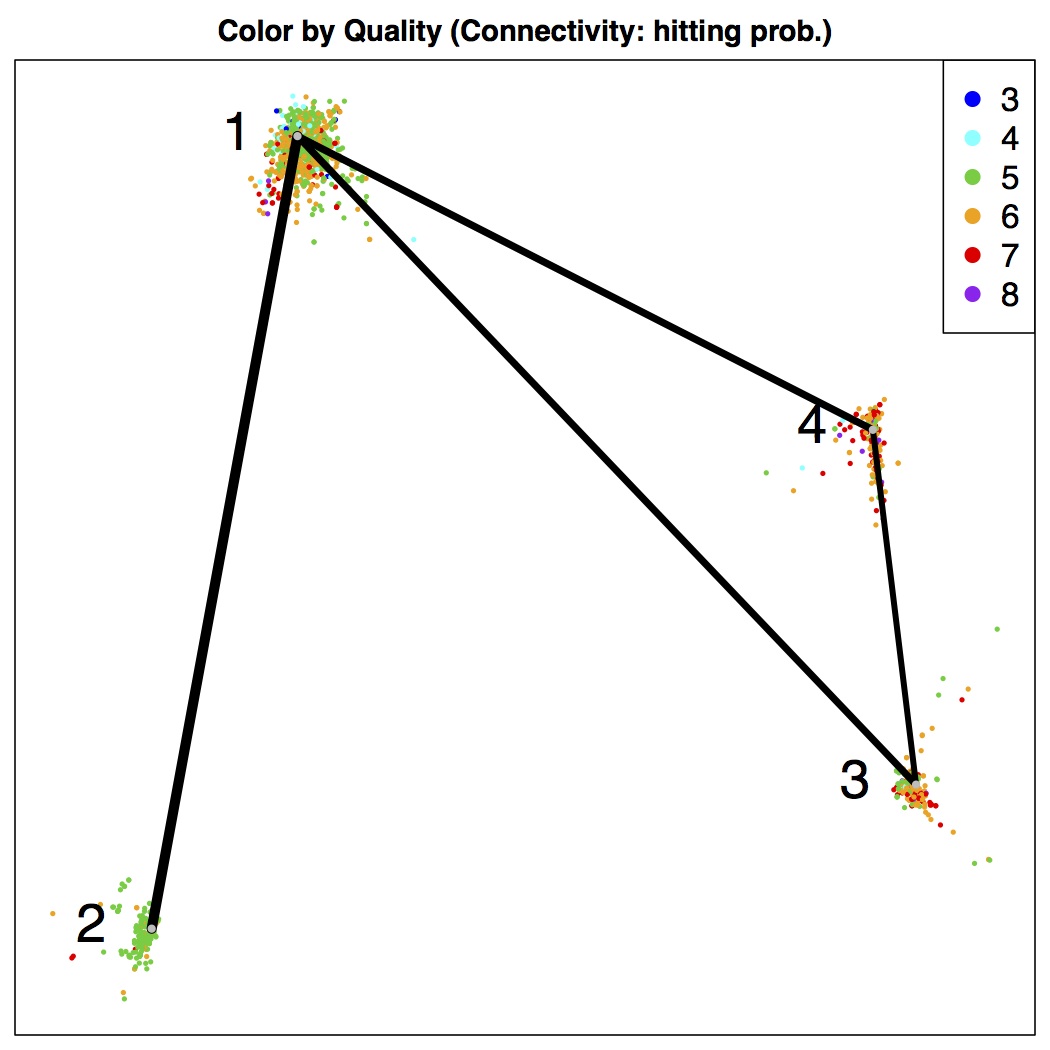} 
	}\\
	\subfigure[Wine quality versus cluster.]
	{
\begin{tabular}{crrrr}
  \hline
 Quality& 1 & 2 & 3 & 4 \\ 
  \hline
  3 &   10 &   0 &   0 &   0 \\ 
  4 &  49 &   0 &   1 &   3 \\ 
  5 & 486 &  135 &  41 & 19 \\ 
  6 & 434 &  25 & 91 &  88 \\ 
  7 &  68 &  3 &  48 &   80 \\ 
  8 &   5 &   0 &   5 &   8 \\ 
   \hline
\end{tabular}
	}\\
	\subfigure[Matrix of connectivity]
	{
\begin{tabular}{rrrrr}
  \hline
 & 1 & 2 & 3 & 4 \\ 
  \hline
1 & -- &\color{red} 0.33 &\color{red} 0.23 &\color{red} 0.23 \\ 
  2 &\color{red} 0.33 & -- & 0.12 & 0.12 \\ 
  3 &\color{red} 0.23 & 0.12 & -- &\color{red} 0.19 \\ 
  4 &\color{red} 0.23 & 0.12 &\color{red} 0.19 & -- \\ 
   \hline
\end{tabular}
	}

\caption{Clustering result for the Wine Quality data ($d=11$). Color denotes
different quality score.
The panel (b) shows the components for each cluster so that we can interpret 
each cluster according to the score distributions.
The first cluster is a normal cluster; the second cluster is a cluster of best wines;
the third cluster is a `better than normal' cluster while the last cluster is a low-score
cluster.
}
\label{Fig::Wine2}
\end{figure}

We apply our methods to the wine quality data set given in 
the UCI machine learning database repository \citep{asuncion2007uci}.
This data set consists of two variants (white and red) of the Portuguese Vinho Verde wine.
The detailed information on this data set is given in \cite{cortez2009modeling}.
In particular, we focus on the red wine, which consists of $n=1599$ observations.
For each wine sample, we have 11 physicochemical measurements:
`fixed acidity',
`volatile acidity',
`citric acid',
`residual sugar',
`chlorides',
`free sulfur dioxide', 
`total sulfur dioxide',
`density',
`pH',
`sulphates' and
`alcohol'.
Thus, the dimension to this dataset is $d=11$.
In addition to the 11physicochemical attributes,
we also have one score for each wine sample.
This score is evaluated by a minimum of three sensory assessors
(using blind tastes), which graded the wine in a scale that ranges from
0 (very bad) to 10 (excellent). The final sensory score is given by the
median of these evaluations.

We apply our methods to this dataset using the same reference rules for bandwidth selection
and picking $n_0$.
The SC-plot is given by Figure~\ref{Fig::WineSC};
we notice that the gap occurs at the fourth and fifth clusters
and $n_0=62.06$ successfully separate these clusters.
Note that the first cluster contains $783$ points so that it does 
not appear in the SC-plot.
We measure the connectivity among clusters via the hitting probability method
and visualize the data in Figure~\ref{Fig::Wine2}.
The following parameters are used in this dataset:
$$
h= 0.599, \quad n_0=62.06, \quad \rho_0=5,\quad\omega_0=0.125.
$$

The wine quality data is very noisy since it involves a human-rating scoring procedure.
However, mode clustering suggests that there
is structure.
From the confusion matrix in Figure~\ref{Fig::Wine2} panel (b), 
we find that each cluster can be interpreted in terms of the score distribution.
The first cluster is like a `normal' group of wines.
It is the largest cluster and the score is normally distributed centering at around $5.5$ 
(the score $5$ and $6$ are the majority in this cluster).
The second cluster is the `bad' group of wines;
most of the wines within this cluster have only score $5$.
The third and fourth clusters are good clusters; the overall quality within
both clusters is high 
(especially the fourth clusters).
Remarkably, the second cluster (bad cluster) 
does not connect to the third and fourth cluster (good cluster).
This shows that our connectivity measure captures some
structure.


\subsection{Seed Data}	\label{sec::seeds}

\begin{figure}
\centering
\includegraphics[width=2 in, height=2 in]{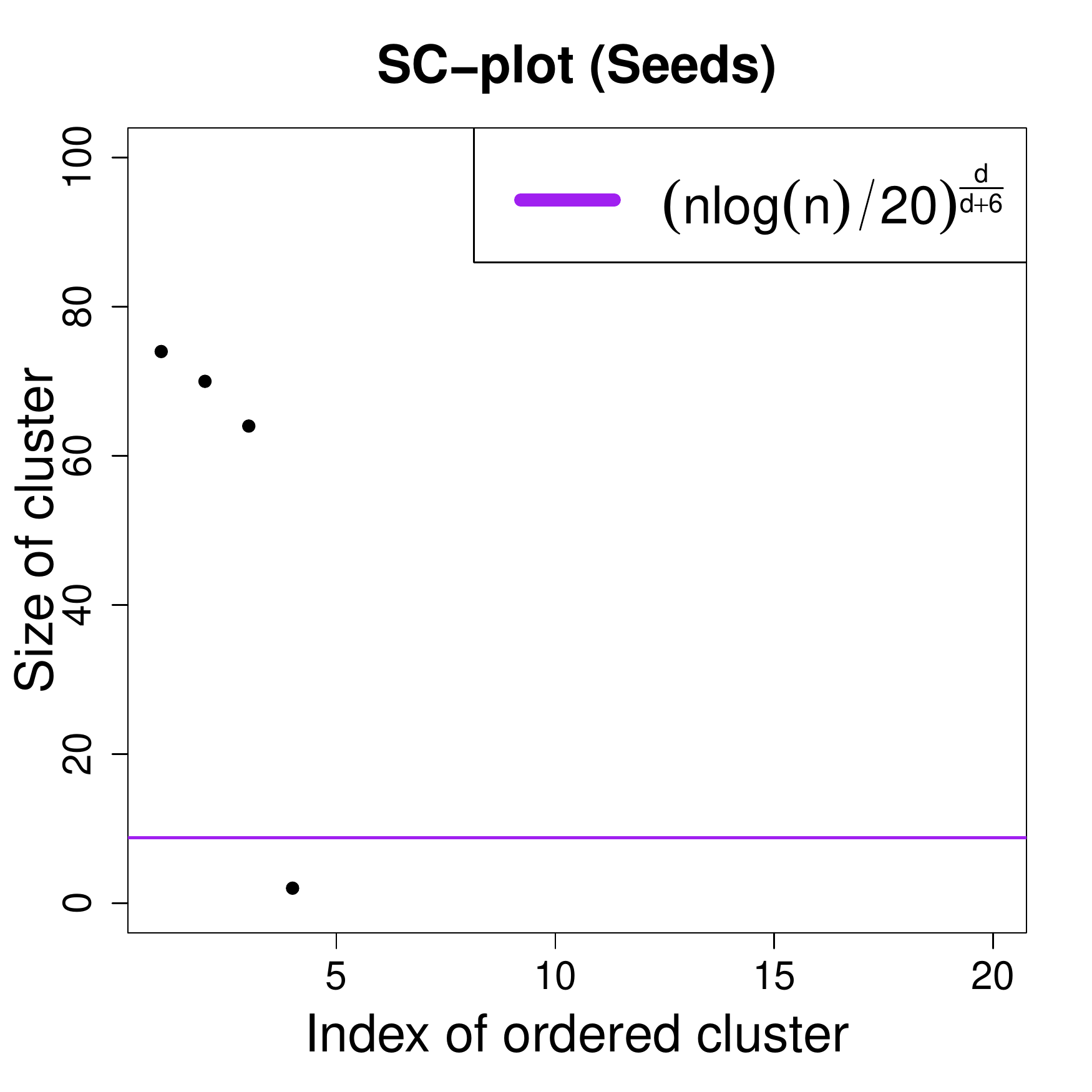} 
\caption{The SC-plot for Seeds data. 
We pick $n_0 = 8.75$ which filters out the fourth small cluster 
(compared to the three large clusters).}
\label{Fig::SeedsSC}
\end{figure}

\begin{figure}
\centering
	\subfigure[Visualization]
	{
		\includegraphics[width=2 in]{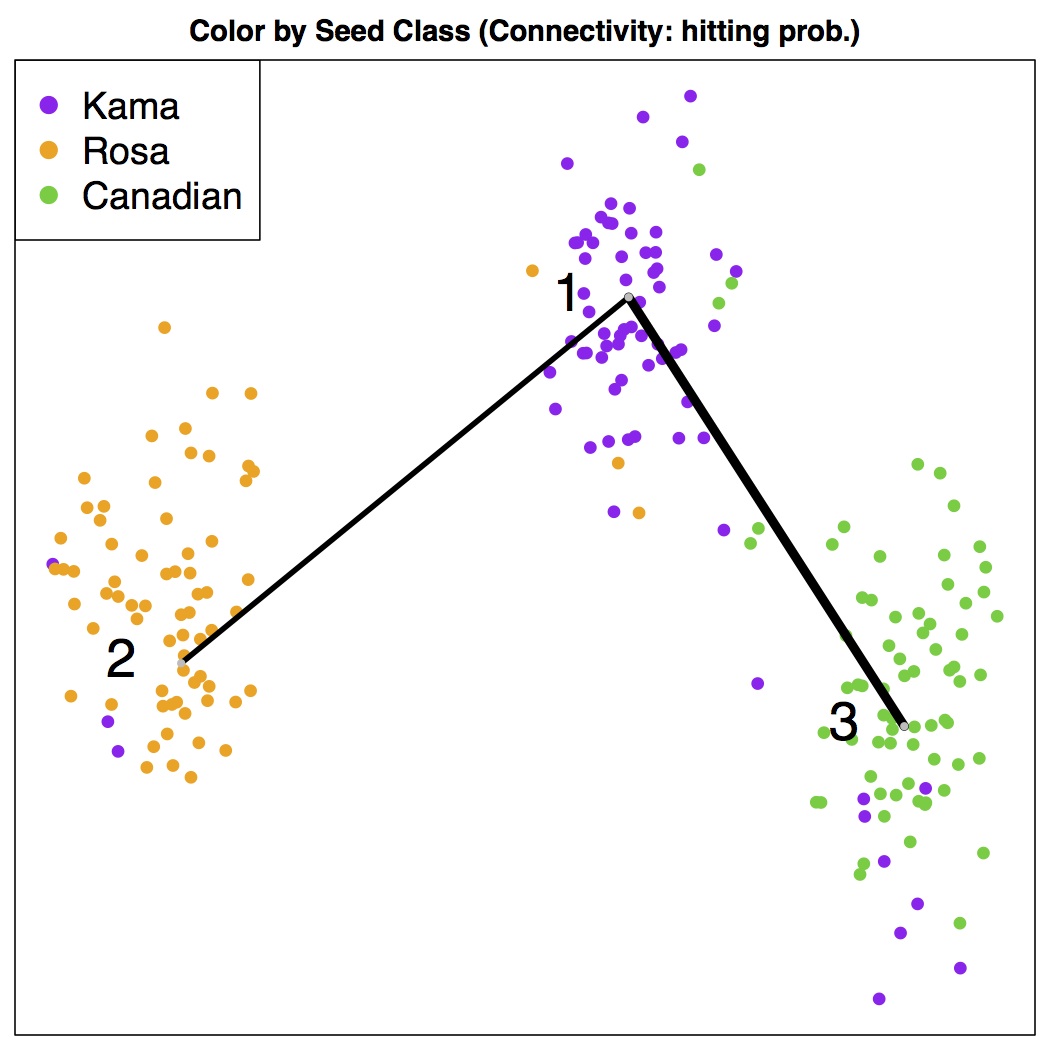} 
	}\\
	\subfigure[Seeds class versus cluster.]
	{
\begin{tabular}{crrr}
  \hline
Class & 1 & 2 & 3 \\ 
  \hline
Kama &   58 &   3 &  9 \\ 
  Rosa &   3 &  67 &   0 \\ 
  Canadian &  3 &   0 &   67 \\ 
   \hline
\end{tabular}
	}\\
	\subfigure[Matrix of connectivity]
	{
\begin{tabular}{rrrr}
  \hline
 & 1 & 2 & 3 \\ 
  \hline
1 & -- & \color{red}0.18 &\color{red} 0.30 \\ 
  2 &  \color{red}0.18 & -- & 0.09 \\ 
  3 & \color{red}0.30 & 0.09 &-- \\ 
   \hline
\end{tabular}
	}

\caption{Clustering result for the Seed data ($d=7$). Color denotes
different classes of seeds. The three clusters represent three classes of
seeds. The fact that the some seeds appear in the wrong cluster
is captured by the connectivity measure (the high connection
between 1-2 and 1-3).
}
\label{Fig::Seed2}
\end{figure}

We apply our methods to the seed data from
the UCI machine learning database repository \citep{asuncion2007uci}.
The seed data is contributed by the authors of \cite{charytanowicz2010complete}.
Some preliminary analysis using mode clustering (mean shift) and 
K-means clustering can be found in \cite{charytanowicz2010complete}.
Scientists examine the kernels from three different variety of wheat: 
`Kama', `Rosa' and `Canadian'; each type of wheat with a randomly selected $70$ sample.
For each sample, a soft X-ray technique is conducted to obtain an $13\times18 $ cm image.
According to the image, we have 7 attributes ($d=7$): `area', `perimeter', `compactness',
`length of kernel', `width of kernel', `asymmetry coefficient' and `length of kernel groove'.

We first normalize each attribute and then perform mode clustering
according to the bandwidth selected by Silverman's rule \eqref{eq::h0}.
We pick $n_0=8.75$ which is reasonable compared with the SC-plot (Figure~\ref{Fig::SeedsSC}).
The visualization, confusion matrix and the matrix of connectivity is given in Figure
\ref{Fig::Seed2}. 
The following
parameters are used in the seeds data
$$
h= 0.613, \quad n_0=8.75, \quad \rho_0=5,\quad\omega_0=0.167.
$$

As can be seen from Figure \ref{Fig::Seed2}, the three clusters
successfully separate the three classes of seeds with little error.
The connectivity matrix in panel (c) explains
the errors in terms of overlapping of clusters.
Some seeds of class `Kama' (corresponding to the third cluster) are 
in the domain of first and second clusters and we see a higher connectivity
among cluster pair 1-2 and 1-3.

\subsection{Comparisons}	\label{sec::comp}
Finally, we compare mode clustering to 
k-means clustering, spectral clustering and hierarchical clustering
for the four real datasets mentioned previously
(Olive Oil, Bank Authentication, Wine Quality and
Seeds).
For the other three methods,
we pick the number of clusters
as the number of significant clusters by mode clustering.
We use a Gaussian kernel for spectral clustering
and complete linkage for hierarchical clustering.
To compare the quality of clustering, 
we use the adjusted Rand index 
\citep{rand1971objective,hubert1985comparing, vinh2009information}.
The result is given in Table \ref{tab::comp}.
A higher adjusted rand index indicates a better match clustering result.
Note that the adjusted rand index may be negative 
(e.g. wine quality dataset for spectral clustering
and hierarchical clustering).
If a negative value occurs, this means that the clustering result
is worse than randomly partitioning the data.
i.e. the clustering is no better than random guessing.

\begin{figure}
\scriptsize
\centering
\begin{tabular}{|l |c|c|c|c|}
\hline
Dataset/Method      & Mode clustering & k-means & Spectral clustering & Hierarchical clustering \\ \hline
Olive Oil           & \bf0.826           & 0.793   & 0.627               & 0.621                   \\ \hline
Bank Authentication &\bf 0.559           & 0.212   & 0.468               & 0.062                   \\ \hline
Wine Quality        &\bf 0.074           & 0.034   & -0.002              & -0.017                  \\ \hline
Seeds               & 0.765           &\bf 0.773   & 0.732               &  0.686                       \\ \hline
\end{tabular}
\label{tab::comp}
\caption{Adjusted rand index for each method. 
Note that the spectral clustering outputs a random result each time due to its implicitly uses
of k-means clustering. Here we only display one instance.
}
\end{figure}

From Figure \ref{tab::comp}, we find that the mode clustering is
the best method for the olive oil data, 
bank authentication dataset, and the wine quality dataset.
For the case that mode clustering is suboptimal, 
the result is still not to far away from the optimal method.
On the contrary, k-means is a disaster for the bank authentication dataset
and is just a little bit better than mode clustering in the seeds dataset.
For the spectral clustering, overall its performance is very good but 
it fails in the wine quality dataset.
The wine quality dataset (Section \ref{sec::WQ}) is known to be
extremely noisy; this might be the reason why every approach does not
give a good result.
However, even the noise level is so huge,
the mode clustering still detect some hidden structures.
See Section \ref{sec::WQ} for more involved discussion.

\section{Conclusion}	\label{sec::conclude}

In this paper, 
we present enhancements to mode clustering methods,
including soft mode clustering,
a measure of cluster connectivity,
a rule for selecting bandwidth,
a method for denoising small clusters,
and new visualization methods for high-dimensional data.
We also establish a `standard procedure' for mode clustering analysis
in Figure~\ref{Fig::flowchart} that can be used to understand the structure
of data even in high dimensions.
We apply the standard procedure to several examples.
The cluster connectivity and visualization methods
apply to other clustering methods as well.

\section*{Acknowledgement}
Yen-Chi Chen is supported by DOE grant number DE-FOA-0000918.
Christopher R. Genovese is supported by DOE grant number DE-FOA-0000918
and NSF grant number DMS-1208354.
Larry Wasserman is supported by NSF grant number DMS-1208354.

\appendix

\section{Appendix: Mixture-Based Soft Clustering}

The assignment vector $a(x)$ derived from a mixture model need not be well defined
because a density $p$ can have many different mixture representations
that can in turn result in distinct soft cluster assignments.

Consider a mixture density
\begin{equation}
p(x) = \sum_{j=1}^{k} \pi_j \phi(x;\mu_j,\Sigma_j)
\label{eq::IDM1}
\end{equation}
where each $\phi(x;\mu_j,\Sigma_j)$ is a Gaussian density function with mean
$\mu_j$ and covariance matrix $\Sigma_j$ and $0\le\pi_j\le 1$ is the
mixture proportion for the $j$-th density such that $\sum_j
\pi_j=1$. 
Recall the latent variable representation of
$p$. Let $Z$ be a discrete random variable such that
\begin{equation}
P(Z=j)=\pi_j, \quad j=1,\cdots,k
\end{equation}
and let
$X|Z \sim \phi(x;\mu_Z,\Sigma_Z)$. Then, consistent with \eqref{eq::IDM1},
the unconditional density for $X$ is
\begin{equation}
p(x) = \sum_z p(x|z)p(z)=\sum_{j=1}\pi_j \phi(x;\mu_j,\Sigma_j)
\end{equation}
It follows that
\begin{equation}
P(Z=j|x) =\frac{\pi_j p(x|Z=j)}{\sum_{s=1}\pi_sp(x|z=s)} =
\frac{\pi_j \phi(x;\mu_j,\Sigma_j)}{\sum_{s=1}\pi_s \phi(x;\mu_s,\Sigma_s)},
\end{equation}
with soft cluster assignment
$a(x) = (a_1(x),\cdots,a_k(x)) =(p(z=1|x),\cdots,p(z=k|x))$.
Of course, $a(x)$ can be estimated from the data
by estimating the parameters of the mixture model.

We claim that the $a(x)$ is not well-defined. 
Consider the following
example in one dimension. Let
\begin{equation}
p(x) = \frac{1}{2} \phi(x;-3,1)+\frac{1}{2} \phi(x;3,1).
\label{eq::IDM2}
\end{equation}
Then by definition 
\begin{equation}
a_1(x) = P(Z=1|x) = \frac{ \frac{1}{2} \phi(x;-3,1)}{\frac{1}{2} \phi(x;-3,1)+\frac{1}{2} \phi(x;3,1)}.
\label{eq::IDM3}
\end{equation}
However, we can introduce a different latent variable representation for $p(x)$ as follows. Let us define
\begin{equation}
p_1(x) =\frac{p(x)1(x\leq 4)}{\int p(x)1(x\leq 4)dx}
\end{equation}
and 
\begin{equation}
p_2(x) =\frac{p(x)1(x> 4)}{\int p(x)1(x> 4)dx}
\end{equation}
and note that 
\begin{equation}
p(x) = \pi p_1(x)+(1-\pi)p_2(x)
\end{equation}
where $\pi = \int p(x)1(x\leq 4)dx$. 
Here, $1(E)$ is the indicator
function for $E$. Let $W$ be a discrete random variable such that
$P(W=1) = \pi$ and $P(W=2) = 1-\pi$ and let $X|W$ has density
$p_W(x)$. Then we have $p(x)= \sum_{w} p(x|w)P(W=w)$ which is the same
density as \eqref{eq::IDM2}. 
This defined the soft clustering
assignment $a(x) = (P(W=1|x),\cdots,P(W=k|x))$ where
\begin{align}
a_1(x) = P(W=1|x) = 1(x\leq 4)
\end{align} 
which is completely different from~\eqref{eq::IDM3}. In fact, for any
set $A\subset \R$, there exists a latent representation of
$p(x)$ such that $a_1(x) = I(x\in A)$. There are infinitely many
latent variable representations for any density, each leading to a
different soft clustering.
The mixture-based soft clustering thus depends on the arbitrary, chosen representation.

\section{Appendix: Proofs}
\begin{proof}[ of Theorem \ref{thm::LM}]

For two vector-value functions $f(x),g(x)\in\mathbb{R}^d$ and two matrix-value functions $A(x), B(x)\in\mathbb{R}^{d_1\times d_2}$,
we define the $\cL^{\infty}$ norms
\begin{equation}
\norm{f-g}_{\max,\infty} = \sup_x \norm{f(x)-g(x)}_{\max},\qquad \norm{A-B}_{\max, \infty} = \sup_x \norm{A(x)-B(x)}_{\max},
\end{equation}
where $\norm{f(x)-g(x)}_{\max}, \norm{A(x)-B(x)}_{\max}$ are the elementwise
maximal norm.
Similarly, for two scalar-value functions $p(x),q(x)$, $\norm{p-q}_{\infty} = \sup_x |p(x)-q(x)|$
is the ordinary $\cL^{\infty}$ norm.

\vspace{0.3 in}
{\bf Modal consistency:}
Our proof consists of three steps. 
First, we show that when $p,\hat{p}_n$ are sufficiently close, 
each local modes $m_j$ corresponds to a unique 
$\hat{m}_j$.
Second, we show that when $\norm{\nabla p- \nabla \hat{p}_n}_{\max,\infty}$ 
and $\norm{\nabla \nabla p-\nabla\nabla \hat{p}_n}_{\max,\infty}$
are small, all the estimated local mode must be near to some local modes.
The first two steps and (M2) construct a condition for
an unique 1-1 correspondence between elements of $\mathcal{M}$ and 
$\hat{\mathcal{M}}_n$.
The last step is to apply Talagrand's inequality to 
get the exponential bound for the probability of the desire condition.

{\bf Step 1:}
WLOG, we consider a local mode $m_j$. 
Now we consider the set 
$$
S_j = m_j\oplus \frac{\lambda_*}{2dC_3}.
$$
Since the third derivative of $p$ is bounded by $C_3$,
$$
\sup_{x\in S_j}\norm{\nabla \nabla p(m_j) - \nabla \nabla p(x)}_{\max}\leq 
\frac{\lambda_*}{2dC_3}\times C_3 
= \frac{\lambda_*}{2d}.
$$
Thus, by Weyl's theorem (Theorem 4.3.1 in \cite{Horn2013}) and condition (M1),
the first eigenvalue is bounded by
\begin{equation}
\sup_{x\in S_j} \lambda_1(x) \leq \lambda_1(m_j)+ d\times\frac{\lambda_*}{2d} \leq -\frac{\lambda_*}{2}.
\label{eq::pf1_1}
\end{equation}
Note that eigenvalues at local modes are negative.
Since $\nabla p(m_j)=0$ and the eigenvalues are bounded around $m_j$,
the density at the boundary of $S_j$ must be less than
$$
\sup_{x\in \partial S_j} p(x) \leq p(m_j)-\frac{1}{2}\frac{\lambda_*}{2}\left(\frac{\lambda_*}{2dC_3}\right)^2
=p(m_j)- \frac{\lambda_*^3}{16d^2C_3^2},
$$
where $\partial S_j = \{x: \norm{x-m_j}= \frac{\lambda_*}{2dC_3}\}$ is the boundary
of $S_j$.
Thus, whenever 
\begin{equation}
\norm{\hat{p}_n-p}_{\infty}< \frac{\lambda_*^3}{16d^2C_3^2},
\label{eq::pf1_2}
\end{equation}
there must be at least one estimated local mode $\hat{m}_j$ within $S_j = m\oplus \frac{\lambda_*}{2dC_3}$.
Note that this can be generalized to each $j=1,\cdots, k$.

{\bf Step 2:}
It is straightforward to see that whenever
\begin{equation}
\begin{aligned}
	\norm{\nabla \hat{p}_n-\nabla p}_{\max,\infty}&\leq \eta_1,\\
	\norm{\nabla\nabla \hat{p}_n-\nabla \nabla p}_{\max,\infty} &\leq  \frac{\lambda_*}{4d},
\end{aligned}
\label{eq::pf1_3}
\end{equation}
the estimated local modes
$$
\hat{\mathcal{M}}_n\subset \mathcal{M}\oplus \frac{\lambda_*}{2dC_3}
$$
by using (M2), triangular inequality and again Weyl's theorem for the eigenvalues.

{\bf Step 3:}
By Step 1 and 2, 
$$
\hat{\mathcal{M}}_n\subset \mathcal{M}\oplus \frac{\lambda_*}{2dC_3}
$$
and for each mode $m_j$ there exists at least one estimated mode $\hat{m}_j$
within $S_j = m_j\oplus\frac{\lambda_*}{2dC_3} $.
Now apply \eqref{eq::pf1_1} and second inequality of \eqref{eq::pf1_3}
and triangular inequality, 
we conclude
\begin{equation}
\sup_{x\in S_j} \hat{\lambda}_1(x)  \leq -\frac{\lambda_*}{4},
\label{eq::pf1_4}
\end{equation}
where $\hat{\lambda}_1(x)$ is the first eigenvalue of $\nabla\nabla \hat{p}_n(x)$.
This shows that we cannot have two estimated local modes within each $S_j$.
Thus, each $m_j$ only corresponds to one $\hat{m}_j$ and vice versa by Step 2.
We conclude that a sufficient condition for the number of modes being the same is
the inequality required in \eqref{eq::pf1_2} and \eqref{eq::pf1_3} i.e. we need
\begin{equation}
\begin{aligned}
	\norm{\hat{p}_n-p}_{\infty}&< \frac{\lambda_*^3}{16d^2C_3^2},\\
	\norm{\nabla \hat{p}_n-\nabla p}_{\max,\infty}&\leq \eta_1,\\
	\norm{\nabla\nabla \hat{p}_n-\nabla \nabla p}_{\max,\infty} &\leq  \frac{\lambda_*}{4d}.
\end{aligned}
\label{eq::pf1_5}
\end{equation}

Let $p_h = \mathbb{E}(\hat{p}_n)$ be the smoothed version of the KDE.
It is well-known in nonparametric theory that (see e.g. page 132 in \cite{scott2009multivariate})
\begin{equation}
\begin{aligned}
	\norm{p_h-p}_{\infty}&=O(h^2),\\
	\norm{\nabla p_h-\nabla p}_{\max,\infty}&=O(h^2),\\
	\norm{\nabla\nabla p_h-\nabla \nabla p}_{\max,\infty} &= O(h^2).
\end{aligned}
\label{eq::pf1_5_1}
\end{equation}
Thus, as $h$ is sufficiently small, we have
\begin{equation}
	\norm{p_h-p}_{\infty}< \frac{\lambda_*^3}{32d^2C_3^2},\qquad
	\norm{\nabla p_h-\nabla p}_{\max,\infty}\leq \eta_1/2,\qquad
	\norm{\nabla\nabla p_h-\nabla \nabla p}_{\max,\infty} \leq  \frac{\lambda_*}{8d}.
\label{eq::pf1_5_2}
\end{equation}
Thus, \eqref{eq::pf1_5} holds whenever
\begin{equation}
\begin{aligned}
	\norm{\hat{p}_n-p_h}_{\infty}&< \frac{\lambda_*^3}{32d^2C_3^2},\\
	\norm{\nabla \hat{p}_n-\nabla p_h}_{\max,\infty}&\leq \eta_1/2,\\
	\norm{\nabla\nabla \hat{p}_n-\nabla \nabla p_h}_{\max,\infty} &\leq  \frac{\lambda_*}{8d}
\end{aligned}
\label{eq::pf1_5_3}
\end{equation}
and $h$ is sufficiently small.

Now applying Talagrand's inequality \citep{Talagrand1996, Gine2002} 
(see also equation (90) in Lemma 13 in \cite{Chen2014} for a similar result), 
there exists constants $A_0,A_1,A_2$ and $B_0,B_1,B_2$
such that for $n$ sufficiently large,
\begin{equation}
\begin{aligned}
	\mathbf{P}\left(\norm{\hat{p}_n-p_h}_{\infty}\geq \epsilon\right) &\leq B_0 e^{-A_0 \epsilon n h^{d}},\\
	\mathbf{P}\left(\norm{\nabla \hat{p}_n-\nabla p_h}_{\max,\infty}\geq \epsilon\right) &\leq B_1 e^{-A_1 \epsilon n h^{d+2}},\\
	\mathbf{P}\left(\norm{\nabla\nabla \hat{p}_n-\nabla \nabla p_h}_{\max,\infty} \geq \epsilon\right) &\leq B_2 e^{-A_2 \epsilon n h^{d+4}}.
\end{aligned}
\label{eq::pf1_6}
\end{equation}
Thus, combining \eqref{eq::pf1_5_3} and \eqref{eq::pf1_6}, we conclude that
there exists some constants $A_3, B_3$ such that 
\begin{equation}
\mathbb{P}(\eqref{eq::pf1_5}\mbox{ holds})\geq 1-B_3e^{-A_3 nh^{d+4}}
\end{equation}
when $h$ is sufficiently small.
Since \eqref{eq::pf1_5} holds implies $\hat{k}_n=k$, we conclude
\begin{equation}
\mathbb{P}(\hat{k}_n\neq k)\leq B_3e^{-A_3 nh^{d+4}}
\end{equation}
for some constants $B_3,A_3$ as $h$ is sufficiently small. This proves modal consistency.

\vspace{0.3 in}
{\bf Location convergence:}
For the location convergence, we assume \eqref{eq::pf1_5} holds so that 
$\hat{k}_n=k$ and each local mode is approximating by an unique estimated local mode.
We focus on one local mode $m_j$ and derive the rate of convergence
for $\norm{\hat{m}_j-m_j}$ and then generalized this rate to all the local modes.

By definition, 
$$
\nabla p(m_j) = \nabla \hat{p}_n(\hat{m}_j) = 0.
$$
Thus, by Taylor expansion and the fact that the third derivative of $\hat{p}_n$ is
uniformly bounded,
\begin{equation}
\begin{aligned}
\nabla \hat{p}_n(m_j)& = \nabla \hat{p}_n(\hat{m}_j) - \nabla \hat{p}_n(m_j) \\
& = \nabla \nabla \hat{p}_n(m_j) (\hat{m}_j - m_j)+ o (\norm{\hat{m}_j - m_j}).
\end{aligned}
\end{equation}
Since we assume \eqref{eq::pf1_5}, this implies 
all eigenvalues of $\nabla \nabla \hat{p}_n(m_j)$ are bounded away from $0$
so that $\nabla \nabla \hat{p}_n(m_j)$ is invertible.
Moreover, 
\begin{equation}
\begin{aligned}
\nabla \hat{p}_n(m_j) &= \nabla \hat{p}_n(m_j)- \nabla p(m_j)\\
& = O(h^2) + O_P\left(\sqrt{\frac{1}{nh^{d+2}}}\right)
\end{aligned}
\end{equation}
by the rate of pointwise convergence in nonparametric theory 
(see e.g. page 154 in \cite{scott2009multivariate}). 
Thus, we conclude
\begin{equation}
\norm{\hat{m}_j - m_j} = O(h^2) + O_P\left(\sqrt{\frac{1}{nh^{d+2}}}\right).
\end{equation}
Now applying this rate of convergence to each local mode and
use the fact that
$$
\Haus\left(\hat{\mathcal{M}}_n, \mathcal{M}\right) = \max_{j=1,\cdots,k} \norm{\hat{m}_j-m_j},
$$
we conclude the rate of convergence for estimating the location.

\end{proof}

\bibliographystyle{abbrvnat}

\bibliography{SoftCluster.bib}

---
\end{document}